\documentclass[aps,prb,twocolumn,groupedaddress,showpacs]{revtex4}
\usepackage{graphicx} 
\usepackage{psfrag} 
\usepackage{color}
\usepackage{amsbsy} 
\usepackage{amssymb} 
\usepackage{lineno}
\usepackage{chemarr}

\definecolor{purple}{rgb}{0.65, 0, 1}
\definecolor{orange}{rgb}{1,.5,0}

\begin{document}

\title[First Assembly Times]{First passage times in homogeneous
nucleation and self-assembly}

\author{Romain Yvinec$^{1}$, Maria R. D'Orsogna$^{2}$, Tom Chou$^{3}$}
\affiliation{$^1$Universit\'e
de Lyon, 
CNRS UMR 5208
Universit\'e Lyon 1
Institut Camille Jordan \\
43 Blvd. du 11 novembre 1918
F-69622 Villeurbanne CEDEX
France}
%\affiliation{$^{1}$Dept. of Mathematics, Universit\'{e} de Lyon,
%  F-69622 Villeurbanne, France}
\affiliation{$^{2}$Dept. of Mathematics, CSUN, Los Angeles, CA 91330-8313}
\affiliation{$^{3}$Depts. of Biomathematics and Mathematics, UCLA, Los
  Angeles, CA, 90095, USA}

\date{\today}

\begin{abstract}
\noindent
Motivated by nucleation and molecular aggregation in physical,
chemical and biological settings, we present a thorough analysis of
the general problem of stochastic self-assembly of a fixed number of
identical particles in a finite volume. We derive the Backward
Kolmogorov equation (BKE) for the cluster probability
distribution. From the BKE we study the distribution of times it takes
for a single maximal cluster to be completed, starting from any
initial particle configuration. In the limits of slow and fast
self-assembly, we develop analytical approaches to calculate the mean
cluster formation time and to estimate the first assembly time
distribution.  We find, both analytically and numerically, that faster
detachment can lead to a {\it shorter} mean time to first completion
of a maximum-sized cluster. This unexpected effect arises from a
redistribution of trajectory weights such that upon increasing the
detachment rate, paths that take a shorter time to complete a cluster
become more likely.
\end{abstract}


\maketitle

\section{Introduction} 
\label{S:intro}

\noindent
The self-assembly of macromolecules and particles is a fundamental
processes in many physical and chemical systems.
%Nucleation and
%molecular aggregation are key processes in numerous biological
%systems, where proteins aggregate to form ion channels, viral capsids,
%and plaques implicated in neurological diseases.  
Although particle nucleation and assembly have been studied for many
decades, interest in this field has recently intensified due to
engineering, biotechnological and imaging advances at the nanoscale
level \cite{WHITESIDES0, WHITESIDES1, GROS}.  Aggregating atoms and
molecules can lead to the design of new materials useful for surface
coatings \cite{SURFACE}, electronics \cite{WIRE}, drug delivery
\cite{LANGER} and catalysis \cite{REBEK}.  Examples include the
self-assembly of DNA structures \cite{DNA0, DNA1} into polyhedral
nanocapsules useful for transporting drugs \cite{BHATIA} or the
self-assembly of semiconducting quantum dots to be used as quantum
computing bits \cite{KROUTVAR}.
                    
Other important examples of molecular self-assembly may be found
in cell physiology or virology where proteins aggregate to form ion
channels, viral capsids and plaques implicated in neurological
diseases.  One example is the rare self-assembly of fibrous protein
aggregates such as $\beta-$amyloid that have long been suspected to
play a role in neurodegenerative conditions such as Alzheimer's,
Parkinson's, and Huntington's disease \cite{SOTO}.  In prion diseases,
individual PrP$^{\rm C}$ proteins misfold into PrP$^{\rm Sc}$ prions
which subsequently self-assemble into fibrils. The aggregation of
misfolded proteins in neurodegenerative diseases is a rare event,
usually involving a very low concentration of prions. Fibril
nucleation also appears to occur slowly; however once a critical size
of about ten proteins is reached, the fibril stabilizes and
the growth process accelerates \cite{NOWAK}.

\begin{figure}[h!]
\begin{center}
\includegraphics[width=3.4in]{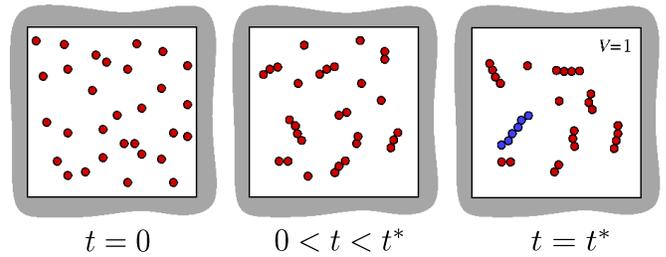}
\caption{Homogeneous self-assembly and growth in a closed unit volume
  initiated with $M=30$ free monomers. At a specific intermediate time
  $0 < t < t^*$
  in this depicted realization, there are six free monomers, four
  dimers, four trimers, and one cluster of size four. For each
  realization of this process, there will be a specific time $t^{*}$
  at which a maximum cluster of size $N=6$ in this example 
  is first formed
  (blue cluster).}
\label{FIG1}
\end{center}
\end{figure}

Viral proteins may also self-assemble to form capsid shells in
the form of helices, icosahedral, dodecahedra, depending on virus
type. A typical assembly process will involve several steps where
dozens of dimers aggregate to form more complex subunits which later
cooperatively assemble into the capsid shell.  Usually, capsid
formation requires hundreds of protein subunits 
that self-assemble over a period of seconds to hours, depending
on experimental conditions \cite{ZLOTNICK, ZLOTNICK2}. 

In addition to these two examples, many other biological processes
involve a fixed ``maximum" cluster size -- of tens or hundreds of
units -- at which the process is completed or beyond which the
dynamics change \cite{OURJCP}.  At times, the assembly process may
involve coagulation and fragmentation of clusters as well, such as in
the case of telomere aggregation in the yeast nucleus
\cite{holcman2012}.  Developing a stochastic self-assembly model with
a fixed ``maximum'' cluster size is thus important for our
understanding of a large class of biological phenomena.

Theoretical models for self-assembly have typically described
mean-field concentrations of clusters of all possible sizes using the
well-studied mass-action, Becker-D\"{o}ring equations
\cite{PENROSE,WATTIS,SMEREKA,OURPRE}.  While Master equations for the
fully stochastic nucleation and growth problem have been derived, and
initial analyses and simulations performed \cite{Kelly,
  BHATT,EBELING}, there has been relatively less work on the
stochastic self-assembly problem. We have recently shown that in
finite systems, where the maximum cluster size is capped, results from
mass-action equations are inaccurate and that in this case a discrete
stochastic treatment is necessary \cite{DLC}.

In our previous examination of equilibrium cluster size distributions
derived from a discrete, stochastic model \cite{DLC}, we found that a
striking finite-size effect arises when the total mass is not
divisible by the maximum cluster size. In particular, we identified
the discreteness of the system as the major source of divergence
between mean-field, mass action equations and the fully stochastic
model.  Moreover, discrepancies between the two approaches are most
apparent in the strong binding limit where monomer detachment is
slow. Before the system reaches equilibrium, or when the detachment is
appreciable, the differences between the mean-field and stochastic
results are qualitatively similar, with only modest quantitative
disparities.

In this paper, we will be interested in the distribution of the first
assembly times towards the completion of a full cluster, which can
only be determined through a fully stochastic treatment.
Specifically, we wish to compute the time it takes for a system of $M$
monomers to first assemble into a complete cluster of size $N$.  We do
not consider coagulation and fragmentation events, but, as a starting
point, focus on attachment and detachment of single monomers.
Statistics of the first assembly time \cite{REDNERBOOK} may shed light
on how frequently fast-growing protein aggregates appear. In
principle, one may also estimate mean self-assembly times starting
from the mean-field, mass action equations, using heuristic arguments.
We will show however that these mean-field estimates yield mean first
assembly times that are quite different from those obtained via exact,
stochastic treatments. 

In the next section, we review the Becker-D\"{o}ring mass-action
equations for self-assembly and motivate the formulation of
approximate expressions for the first assembly time
distributions. These will be shown to be poor estimates of the true
distribution functions, leading us to consider the full stochastic
problem in Section III. Here, we derive the Backward Kolmogorov
equation associated with the self assembly process and illustrate how
to formally solve it through the corresponding eigenvalue
problem. In Section IV, we explore three limits of the stochastic
self-assembly process and derive analytic expressions for the mean
first assembly time in the strong and weak binding limits.  Results
from kinetic Monte-Carlo (KMC) simulations are presented in Section V
and compared with our analytical estimates.  Finally, we discuss the
implications of our results and propose further extensions in the
Summary and Conclusions.

\section{Mass-action model of homogeneous nucleation and self-assembly}

\noindent
The classic mass-action description for spontaneous, homogeneous
self-assembly is the Becker-D\"{o}ring model \cite{BD}, where the
concentrations $c_{k}(t)$ of clusters of size $k$ obey

\begin{eqnarray}
\dot{c}_{1}(t) & = & -p_{1}c_{1}^{2}
- c_{1}\sum_{j=2}^{N-1}p_{j}c_{j}  + 2q_{2} c_{2} +
\sum_{j=3}^{N}q_{j}c_{j} 
\nonumber \\
\dot{c}_{2}(t) & = & \displaystyle  = -p_{2}c_{1}c_{2}+ {\frac {p_{1}}
  2}
c_{1}^{2} - 
q_{2} c_{2} + q_{3}c_{3} \nonumber \\
\dot{c}_{k}(t) & = & -p_{k}c_{1}c_{k}  + p_{k-1}c_{1}c_{k-1} - q_{k} c_{k} 
+q_{k+1} c_{k+1} \nonumber \\
\dot{c}_{N}(t) & = & p_{N-1}c_{1}c_{N-1} -q_{N} c_{N},
\label{HOMOEQN0}
\end{eqnarray}

\noindent 
where $p_{k}$ and $q_k$ are the monomer attachment and
detachment rates to and from a cluster of size $k$.  A typical initial
condition is $c_{k}(t=0) = (M/V)\delta_{k,1}$, representing an initial
state comprised only of free monomers. For simplicity we
set the volume $V=1$. The above equations can be numerically
integrated to find the time-dependent concentrations $c_{k}(t)$ for
any set of attachment and detachment rates.  We have previously shown
that Eqs.~\ref{HOMOEQN0} provide a poor approximation to the expected
number of clusters when the total mass $M$ and the maximum cluster
size $N$ are comparable in magnitude \cite{DLC}.

Although mass action equations provide approximations to mean
concentrations, they do not directly describe any statistical property
of the modeled system. Nonetheless, one may be able to heuristically
derive estimates of quantities such as mean first assembly times.  To
estimate the mean time to completion of the first maximum cluster, we
must consider a truncated set of mass-action equations which treats
maximum clusters as ``absorbing states'' so that once maximum clusters
are formed, the process is stopped and the time recorded.  Thus, we
set $q_N=0$ in Eqs.~\ref{HOMOEQN0} so that once clusters of size $N$
are formed, no detachment is allowed. This choice ensures that
completed assembly events will not influence the dynamics of any of
the remaining smaller clusters.

To estimate the mean first assembly time we may invoke the statistical
concept of survival probabilities, and heuristically combine it with
the deterministic solutions of Eq.~\ref{HOMOEQN0}.  Following standard
notation, we denote by $S(t)$ the probability that the system has not
yet formed a maximal cluster. This quantity is also known as the
``survival'' probability. Its dynamics can be expressed using the
probability flux $J_{N}$ out of the last not fully formed maximal
cluster state, or equivalently into the maximal one,
conditioned on the system still surviving so that

\begin{figure}[t]
\begin{center}
\includegraphics[width=3.35in]{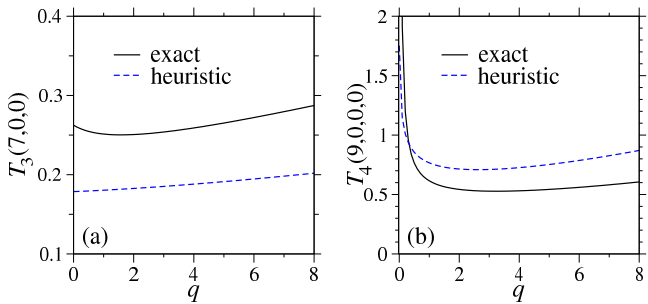}
\caption{Mean first assembly times evaluated via the heuristic
  definition Eq.\,\ref{HEUR} (dashed line) and as a function of the
  detachment rate $q_i = q$, for $M=7$, $N=3$ in panel (a) and for
  $M=9$, $N=4$ in panel (b).  Here $p_i = p = 1$.  We also show the
  exact results (solid line) obtained via the stochastic formulation
  in Eq.\,\ref{analytic} which we derive in Section III.  Qualitative
  and quantitative differences between the two approaches arise, which
  become even more evident for $N>3$, $q \to 0$, as we shall later
  discuss. These discrepancies underline the need for a stochastic
  approach.}
\label{HEUR_N=3}
\end{center}
\end{figure}

\begin{equation}
{{\rm d} S(t) \over {\rm d} t} = -J_{N}(t |\,\mbox{surviving up to time }\,t).
\label{FLU}
\end{equation}

\noindent
The flux $J_{N}(t |\,\mbox{surviving up to time }\,t)$ conditioned on 
survival up to time $t$ is 
not readily found, but a mean-field approximation can be 
applied by assuming $J_{N}(t |\,\mbox{surviving up to time }\,t) \approx 
J_{N}(t) S(t)$, where $J_{N}(t)$ is the unconstrained mean particle
flux. Thus, the mean field approximation for the evolution of
the survival probability becomes

\begin{equation}
{{\rm d}S(t) \over {\rm d} t} \simeq -J_{N}(t)S(t).
\end{equation}

\noindent
To proceed, we may use deterministic 
results for $J_N(t)$ 

\begin{equation}
J_{N}(t) \simeq  p_{N-1}c_{1}(t)c_{N-1}(t),
\end{equation}

\noindent
so that the survival probability can be estimated as

\begin{equation}
\label{EST2}
S(t) = \exp \left[-p_{N-1}\int_{0}^{t}c_{1}(t')c_{N-1}(t') {\rm d}t' \right] =
e^{-c_{N}(t)}.
\end{equation}

\noindent
Note that while Eq.\,\ref{EST2} satisfies $S(t=0) = 1$, $S(t \to \infty)
\not\to 0$, due to 
$c_N(t \to \infty)$ being finite.  As a consequence, the
derived first assembly time will always be infinitely large, since the
system has a finite survival probability even for $t \to \infty$,
making the approximation invalid. Alternatively, we may approximate
Eq.\,\ref{FLU} as

\begin{equation}
{{\rm d}S(t) \over {\rm d}t} = -J_{N}.
\label{SJ}
\end{equation}
This relationship assumes that the system is always is a surviving state 
(not yet formed a maximum-sized cluster). However, Eq.~\ref{SJ} also 
yields unphysical results at long times.  A deterministic
approximation that yields physically reasonable results can be
obtained by finding the time at which the concentration of clusters of
size $N$ reaches unity
\begin{equation}
c_{N}(T_N) \equiv 1,
\label{HEUR}
\end{equation}
%\vspace{3mm}

\noindent and imposing $q_{N}=0$ in Eqs.\,\ref{HOMOEQN0}.  As an example, we
consider the case $M=7$, $N=3$ for $p_i =p= 1, q_{i \neq 3}= q$ (and
$q_3=0$ as illustrated above), find $c_N(T_N)$ from
Eqs.\,\ref{HOMOEQN0}, and plot the mean first assembly time obtained
via Eq.\,\ref{HEUR} in Fig.~\ref{HEUR_N=3}(a).  For completeness we
also show the exact results obtained via the full stochastic treatment
in Eq.~\ref{analytic}, the derivation of which we will focus on below.
What clearly arises from Fig.~\ref{HEUR_N=3}(a) is that while the mean
first assembly times obtained stochastically and via the mean-field
equations are of the same order of magnitude, they are also quite
different and show even qualitative discrepancies. For example, the
stochastic mean first assembly time is non-monotonic in $q$, while the
simple mean-field estimate is an increasing function of $q$.
Discrepancies between the heuristic and exact stochastic results exist
also for the case $M=9$, $N=4$ shown in Fig.\,\ref{HEUR_N=3}(b).
Here, most notably we can point out that for $q=0$, while the exact
mean first assembly time calculated according to our stochastic
formulation diverges, it remains finite in the heuristic derivation.
We shall later see that this trend will persist for all choices $N >
3$ and that the heuristic approach does not yield accurate
estimates. A stochastic treatment is thus necessary and is the subject
of the remainder of this paper.

\section{Backward Kolmogorov Equation}

\noindent To formally derive first assembly times for our nucleation
and growth process it is necessary to develop a discrete, stochastic
treatment.  We thus define $P(n_{1}, n_{2}, \ldots, n_{N}; t\vert
m_{1}, m_{2}, \ldots, m_{N};0)$ as the probability that the system
contains $n_{1}$ monomers, $n_{2}$ dimers, $n_{3}$ trimers, etc, at
time $t$, given that the system started from a given initial
configuration $(m_{1}, m_{2}, \ldots m_{N})$ at $t=0$. In this
representation, the Forward Master equation corresponding to
self-assembly with exponentially-distributed monomer binding and
unbinding events is given by \cite{DLC}

\begin{widetext}
\begin{eqnarray}
\nonumber
\dot{P}(\{n\};t\vert \{m\}, 0) &= &
-\Lambda(\{n\})P(\{n\};t\vert \{m\}, 0) +\frac {p_{1}}
  {2}(n_{1}+2)(n_{1}+1)W^{+}_{1}W^{+}_{1}W^{-}_{2}P(\{n\};t\vert \{m\},
0) \\
\nonumber
&& + q_{2}
(n_{2}+1)W^{+}_{2}W^{-}_{1}W^{-}_{1}P(\{n\};t\vert \{m\}, 0) +
\sum_{i=2}^{N-1}p_{i}(n_{1}+1)(n_{i}+1)W^{+}_{1}W^{+}_{i}W^{-}_{i+1}P(\{n\};t\vert
\{m\}, 0) \\
&&  + \sum_{i=3}^{N}q_{i}
(n_{i}+1)W^{-}_{1}W^{-}_{i-1}W^{+}_{i}P(\{n\};t\vert \{m\}, 0),
\label{MASTER}
\end{eqnarray}
\end{widetext}

\noindent
where $P(\{n\},t) = 0$ if any $n_{i} < 0$ and

\begin{equation}
\Lambda(\{n\}) = {\frac{p_{1}} 2} n_{1}(n_{1}-1) + \sum_{i=2}^{N-1}\!
p_{i}n_{1}n_{i} + \sum_{i=2}^{N}q_{i}n_{i}, \nonumber
\end{equation}

\noindent
is the total rate out of configuration $\{n\}$. Here, $W^{\pm}_{j}$
are the unit raising/lowering operators on the number of clusters of
size $i$ so that

\begin{eqnarray*}
&& W^{+}_{1}W^{+}_{i}W^{-}_{i+1}P(\{n\};t\vert \{m\};0) \equiv \\
\nonumber
&&
\hspace{1cm}
P(n_{1}+1,\ldots,n_{i}+1,n_{i+1}-1,\ldots;t\vert \{m\};0). \nonumber
\end{eqnarray*}

\noindent
The Master equation can be written in the form 
$\dot {\bf P} = {\bf A}{\bf P}$, where ${\bf P}$ is the vector of the
probabilities of all possible configurations and ${\bf A}$ is the
matrix of transition rates between them.
%
%To accurately compute entire assembly time distributions, particularly
%for small particle numbers $M$, it is convenient to consider the
%state-space shown in Fig.~\ref{TREE}, where we consider the explicit
%cases $M=7$ and $N=3$, and $M=8$ and $N=3$.
%
%We wish to evaluate the first time it takes for the system to reach
%any ``absorbing'' state -- one where a cluster of maximal size $N$ is
%fully assembled -- having started from a given initial configuration.
%For example, if $N=3$, absorbing states are those where $n_{N=3} \geq
%1$.  
%The arrival time from a given initial configuration to any
%absorbing state will depend on the specific trajectory taken by the
%system. Upon averaging these arrival times over all paths starting
%from the initial configuration $\{ m\}$ and ending at any absorbing
%state, weighted by their likelihood, we can find the overall
%probability distribution of the time it takes to first assemble a
%complete cluster of size $N$.
%
The natural way of computing the distribution of first assembly times
is to consider the Backward Kolmogorov equation (BKE) describing the
evolution of $P(n_{1}, n_{2}, \ldots, n_{N}; t\vert m_{1}, m_{2},
\ldots, m_{N};0)$ as a function of local changes from the initial
configuration $\{m\}$. The BKE can be expressed as

\begin{figure}[t]
\begin{center}
\includegraphics[width=3.4in]{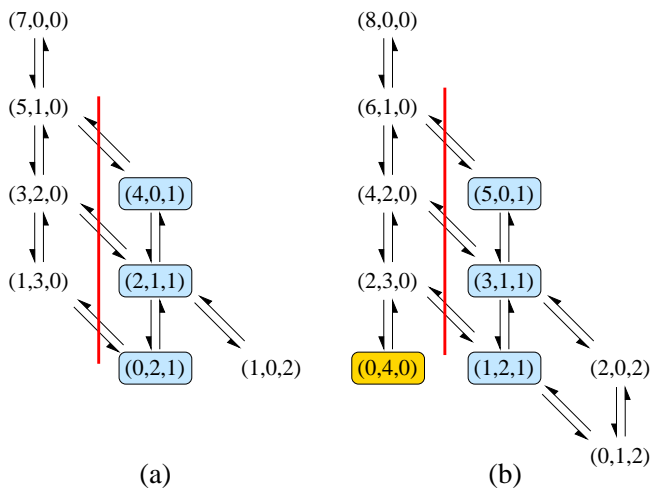}
\caption{Allowed transitions in stochastic self-assembly starting from
  an all-monomer initial condition. In this simple example, the
  maximum cluster size $N=3$.  (a) Allowed transitions for a system
  with $M=7$. Since we are interested in the first maximum cluster
  assembly time, states with $n_{3}=1$ constitute absorbing
  states. The process is stopped once the system crosses the vertical
  red line. (b) Allowable transitions when $M=8$.  Note that if
  monomer detachment is prohibited ($q=0$), the configuration
  $(0,4,0)$ (yellow) is a trapped state. Since a finite number of
  trajectories will arrive at this trapped state and never reach a
  state where $n_{3}=1$, the mean first assembly time $T_{3}(8,0,0)
  \to \infty$ when $q=0$.} 
\label{TREE}
\end{center}
\end{figure}

\begin{widetext}
\begin{eqnarray}
\nonumber
 \dot{P}(\{n\};t\vert \{m\}, 0) &= & 
-\Lambda(\{m\})P(\{n\};t\vert \{m\};0)+ 
\frac{p_{1}}{2} m_{1}(m_{1}-1)W^{+}_{2}W^{-}_{1}W^{-}_{1}
P(\{n\};t\vert \{m\};0) 
\\
\nonumber
&& 
+q_{2}m_{2}W^{-}_{2}W^{+}_{1}W^{+}_{1}
P(\{n\};t\vert \{m\};0)+
\sum_{i=2}^{N-1}p_{i}m_{1}m_{i}W^{-}_{1}W^{-}_{i}W^{+}_{i+1}P(\{n\};t\vert\{m\};0) \\
\label{MASTERBACK}
&& + 
\sum_{i=3}^{N}q_{i}m_{i}W^{+}_{1}W^{+}_{i-1}W^{-}_{i}P(\{n\};t\vert
\{m\};0),
\end{eqnarray}
\end{widetext}

\noindent 
where the operators $W_{i}^{\pm}$ act on the initial configuration
index $m_{i}$.  
%It is straightforward to verify that
%Eq.~\ref{MASTERBACK} is the adjoint of Eq.~\ref{MASTER}.  
In the vector representation, the BKE is $\dot {\bf P} = {\bf
  A}^{\dagger}{\bf P}$, where ${\bf A}^{\dagger}$ is the adjoint of
the transition matrix ${\bf A}$ as can be verified by comparing
Eqs.\,\ref{MASTER} and \ref{MASTERBACK}.  The utility of using the BKE
is that Eq.\,\ref{MASTERBACK} can be used to determine the evolution
of the survival probability, that can be naturally defined as

\begin{equation}
\label{summ}
S(\{m\};t) \equiv \sum_{\{n\},n_{N}=0}P(\{n\};t\vert \{m\};0),
\end{equation}

\noindent
where we have made explicit the dependence on the initial
configuration $\{m\}$.  In Eq.\,\ref{summ} the sum is restricted to
configurations where $n_{N}=0$ so as to include only ``surviving''
states that have not yet reached any of the ones where $n_N \geq 1$.
$S(\{m\};t)$ thus describes the probability that no maximum cluster
has yet been formed up to time $t$, given that the system started in
the configuration $\{ m\}$ at $t=0$.  One can now similarly sum
Eq.~\ref{MASTERBACK} over all final states with fixed $n_N=0$ to find
that $S(\{m\};t)$ also obeys Eq.\,\ref{MASTERBACK} with
$P(\{n\};t\vert \{m\}, 0)$ replaced by $S(\{m\};t)$, along with the
definition $S(m_{1}, m_{2},\ldots, m_{N}\geq 1; t) = 0$ and the
initial condition $S(m_{1}, m_{2},\ldots, m_{N}=0; 0) = 1$. In the
vector representation where each element of ${\bf S}$ corresponds to a
particular initial configuration, the general evolution equation for
the survival probability is $\dot{\bf S} = {\bf A}^{\dagger}{\bf S}$,
where we consider only the subspace of ${\bf A}{^\dagger}$ on
nonabsorbing states.  Solving the matrix equation for ${\bf S}$ leads
to a vector of first assembly time distributions

\begin{equation}
\label{gi}
{\bf G}
\equiv -{\frac{\partial{\bf S}}
{\partial t}},
\end{equation}

where each element of ${\bf G}$ represents the first assembly time
distribution starting from a different initial cluster
configuration. Appendix \ref{APPENDIXA} explicitly details the calculation
procedures required to compute ${\bf S}$, ${\bf G}$, and the moments
of the first assembly times. For example, using Eq.~\ref{timeone}, 
we find 

%For example, for $N=3$, $M=7$ and assuming constant
%$p_i=p$ and $q_i=q$, we find

\begin{eqnarray}
\hspace{-0.8cm}
T_{3}(7,0,0) &=& \frac{1}{105 p^2}\frac{744 p^3+487 p^2 q +60 p q^2+2q^3}
{27 p^2+20p q+2q^2}, 
\label{analytic}
\end{eqnarray}

%\hspace{-0.8cm}
%T_{3}(5,1,0) &=& \frac{1}{105 p^2}\frac{609 p^3 +387p^2 q+50 p q^2+2q^3}
%{27 p^2+20 p q+2q^2}, \\
%\hspace{-0.8cm}
%T_{3}(3,2,0) &=& \frac{1}{105 p^2}\frac{630 p^3 +357p ^2 q+44 p q^2+2q^3}
%{27 p^2+20 p q+2q^2}, \\
%\hspace{-0.8cm}
%T_{3}(1,3,0) &=& \frac{1}{105 p^2}\frac{945 p^3 +385q p^2+ 42 p q^2+2q^3}
%{27 p^2+20 p q+2q^2}. 
%\label{analytic-end}
%\end{eqnarray}

\noindent
where we have assumed $N=3$, $M=7$, and $p_{i} = p$, $q_{i}=q$ are
constants.  The label $(7,0,0)$ indicates an initial condition
consisting of $M=7$ monomers, no dimers, and no trimers. Corresponding
expressions for the mean first assembly time arise for different
initial conditions, such as {\it e.g.}, $(5,1,0)$, $(3,2,0)$, or
$(1,3,0)$. These exact expressions for the mean first assembly times 
are non-monotonic in both $q$ and $p$,
indicating that there are optimal $q/p$ ratios for which the first
assembly times are smallest. We will discuss the monotonicity of
$T_{N}(\{ m\})$ below, both in the limit of fast and slow detachment.
For simplicity, we will retain the assumption of uniform $p_i=p$ and
$q_i=q$ throughout the remainder of this work and henceforth rescale
time in units of $p^{-1}$.  With this choice, $q \gg 1$ represents
fast detachment, while $q \ll 1$ represents slow detachment.
$T_{3}(7,0,0)$ has already been plotted in Fig.\,\ref{HEUR_N=3}(a),
contrasting it against the heuristic approximation of
Eq.\,\ref{HEUR}. A similar matrix approach can be used for the case
$M=9$, $N=4$ yielding a cumbersome but exact expression for
$T_4(9,0,0,0)$ that is plotted in Fig.\,\ref{HEUR_N=3}(b).

\section{Results and Analysis}

\noindent In this section we study the properties of the first
assembly time in the irreversible detachment limit, when $q=0$, and in
the limits of slow ($0< q \ll 1$) and fast detachment ($q \gg 1$).

\subsection{Irreversible limit ($q=0$)}
\noindent
First consider $N=3$ and irreversible self-assembly where $q=0$.  
In this case, the matrix ${\bf A}^{\dagger}$ is bidiagonal and 
the analysis outlined in Appendix \ref{APPENDIXB} yields 
the exact expression for any starting configuration:

\begin{widetext}
\begin{eqnarray}
T_3(M-2n,n,0) = \frac{2}{(M-2n)(M-1)}
\left[1 + \sum_{j=1}^{[ M/ 2]}
  \prod_{k=n+1}^j
\frac{(M-2k +2) (M-2k +1)}{(M-2k)(M-1)}
\right].
\label{TN3}
\end{eqnarray}
\end{widetext}

\noindent
Note that when $q=0$ the mean first assembly time is finite when $M$
is odd, but is infinite if $M$ is even.  This can be understood from
the example $M=8$, $N=3$ illustrated in Fig.\,\ref{TREE}(b), where a
``trapped'' state arises. In this case, there is a finite probability
that the system arrives in the state $(0,4,0)$ trapping it there
forever since the assembly process is irreversible and detachment
would be the only way out. Therefore, averaging over trajectories that
include these ``traps'', the mean assembly time will be infinite.  For
$q=0$, we can show that a trapped state exists for any $M$ and $N \geq
4$, yielding infinite assembly times.  A trapped state arises when
all free monomers have been depleted ($n_{1}=0$) before a maximum
cluster has been able to assemble ($n_N =0$).  In this case, the total
mass must be distributed according to

\begin{eqnarray}
\label{decomp}
M = \sum_{j=2}^{N-1} j n_j.
\end{eqnarray}

\noindent
It is not necessarily the case that this decomposition is possible for
all $M$ and $N$, but if it is, then we have a trapped state and the
first assembly time is infinite. To show that the decomposition holds
for $N \geq 4$ and for all $M$, we write $M = \sigma (N-1) +j$ where
$\sigma$ is the integer part $[ M / (N-1)]$, so that $ 1 \leq j \leq
N-2$.  Now, if $j \neq 1$, then the decomposition is achieved with
$n_{N-1} = \sigma$, $n_j=1$, and all other $n_k =0$ for $k \neq j,
(N-1)$. We have thus constructed a possible trapped state.  If instead
$j=1$, then we can rewrite $M = (\sigma-1)(N-1) + (N-2) + 2$ so that
the decomposed state is defined by $n_{N-1} = \sigma -1$, $n_{N-2} =
1$ and $n_{2}=1$, with all other values of $n_k=0$. This proves that
for all $M,N$ there are trapped states for $q=0$.  The only exception
is when $N=3$, where the last decomposition does not hold, since $N-2
=1$ and by definition, monomers are not allowed in trapped states.
Indeed, for $N=3$, Eq.\,\ref{decomp} gives $M = 2 n_2$ as the only
trapped state, which is possible only for $M$ even.  The case $M=7$
and $N=3$ is shown in Fig.\,\ref{TREE}(a).

According to our stochastic treatment, the possibility of trajectories
reaching trapped states for $q=0$ exists for any value of $M,N \geq
4$, giving rise to infinite mean first assembly times. This behavior
is not mirrored in the mean-field approach for $q=0$, where $c_N(T_N)
= 1$ for finite $T_N$ depending on initial conditions, if $M$ is large
enough as can be seen in Fig.\,\ref{HEUR_N=3}(b).  For $N = 4$, $M=9$,
indeed $T_4(9,0,0,0)$ can be evaluated from Eqs.\,\ref{HOMOEQN0} as
$c_4(1.7527) = 1$.  In the irreversible binding limit, we may thus
find instances where the exact stochastic treatment yields infinite
first assembly times due to the presence of traps, while in the
mean-field, mass action case, the mean first assembly time is finite.
This leads us to expect that mean-field approximation to the first
assembly time will be inaccurate when $q > 0$, but small. Here, the 
trapped states (when $q=0$) retain the system for a long time. 

%\subsection{Conditional first assembly times for $q=0$}
%\noindent

Since infinite mean first assembly times are a consequence of the
existence of trapped states one may ask what is the mean first
assembly time \textit{conditioned} on traps not being visited.  To
this end, we explicitly enumerate all paths towards the absorbed
states and average the mean first assembly times only over those that
avoid such traps \cite{Marcus:1968, Lushnikov:1978}.  To be more
concrete, we first consider the case $N=3$. Here, in order to reach
the absorbing state where $n_3=1$, one or more dimers must have
formed. Let us thus consider the specific case $1 \leq n_2 \leq
\left[\frac {M-1} 2\right]$.  Here, the second bound arises because
after $n_2$ dimers have formed, at least one free monomer must remain
in order to attach to one of the $n_2$ dimers to form the first
trimer. Since at every iteration both the formation of a dimer and of
a trimer can occur, the probability of a path that leads to a
configuration of exactly $n_2$ dimers is given by

\begin{eqnarray}
\label{prob1}
\prod_{k=0}^{n_2-1}\frac{(M-2k)(M-2k-1)}{(M-2k)(M-2k-1) + 2 (M-2k)k},
\end{eqnarray}

\noindent
The above quantity must be multiplied by the probability that after
$n_2$ dimerizations, a trimer is formed, which occurs with probability

\begin{eqnarray}
\label{prob2}
\frac {2 n_2 (M-2 n_2)}{(M-2n_2)(M-2n_2 -1) + 2 (M-2 n_2)n_2 }.
\end{eqnarray}

\noindent
Upon simplifying the product of the two probabilities
in Eqs.\,\ref{prob1} and \ref{prob2}, we find that
the probability $W_{n_2}$ for a path where $n_2$ dimers are created
before the final trimer is assembled is given by

\begin{eqnarray*}
%\label{paths}
W_{n_2} = \frac{2 n_2}{(M-1)^{n_2 +1}} \prod_{k=0}^{n_2-1} (M -2k- 1). 
\end{eqnarray*}

\noindent
Note that if $M$ is even, we must discard paths where $2 n_2 = M$,
since, as described above, this case represents a trap with no
monomers to allow for the creation of a trimer. According to
Eq.\,\ref{prob1}, the realization $2 n_2 = M$ occurs with probability

\begin{eqnarray}
\label{W2}
W_{\frac M 2} = \frac{(M-3)!!}{(M-1)^{\frac M 2-1} M}.
\end{eqnarray}

\noindent
Thus for $M$ even, $W_{\frac M 2}$ represents the probability the
system will end in a trap. We must now evaluate the time the system
spends on each of the trap-free paths. Note that the exit time from a
given dimer configuration $(M-2k,k,0)$ is a random variable taken from
an exponential distribution with rate parameter given by the
dimerization rate, $\lambda_{d,k} = (M-2k)(M-2k-1)/2$.  However, the
formation of a trimer is also a possible way out of the dimer
configuration, with rate $\lambda_{t,k} =(M-2k) k$. The time to exit
configuration $(M-2k,k,0)$ thus is itself an exponentially distributed
random variable with rate $\lambda_k$ given by the sum of the two
rates \cite{McQuarrie:1967}

\begin{eqnarray*}
\lambda_{k} = \lambda_{d,k} + \lambda_{t,k} = \frac{(M-2k) (M-1)} 2.
\end{eqnarray*}

\noindent
The typical time out of configuration $(M-2k,k,0)$ is thus given by
$1/\lambda_{k}$. Upon summing over all possible values $0 \leq k \leq
n_2$, we find the typical time for the system to go through $n_2$
dimerizations

\begin{eqnarray*}
T_{n_2} = \sum_{k=0}^{n_2} \frac 1 {\lambda_k} = \sum_{k=0}^{n_2}
\frac{2}{(M-2k) (M-1)}.
\end{eqnarray*}

\noindent
Finally, we can write the mean first assembly time as

\begin{eqnarray}
\label{finalFPT}
T_{3}(M,0,0) = \sum_{n_2=1}^{\left[\frac {M-1} 2 \right]} W_{n_2} T_{n_2}.
\end{eqnarray}

\noindent
It can be verified that for $M$ odd, Eq.\,\ref{finalFPT} is the same
as Eq.\,\ref{TN3}, since the integer part that appears in the sum in
Eq.\,\ref{finalFPT} is the same as its argument, thus including all
paths. For $M$ even, paths with $2 n_2 = M$ are discarded, yielding a
mean first assembly time averaged over trap-free configurations.

Similar calculations can be carried out for larger $N$; however,
keeping track of all possible configurations before any absorbed state
can be reached becomes quickly intractable. For example, when $N=4$
one would need to consider paths with a specific sequence of $n_{2,k}$
dimers formed between the creation of $k$ and $k+1$ trimers until
$n_3$ trimers are formed. The path would be completed by the formation
of a cluster of size $N=4$. We would then need to consider all
possible choices for $1 \leq n_3 \leq \left[ \frac{M-1} 3\right]$ such
that traps are avoided and evaluate the typical time spent on each
viable path. Because of the many branching possibilities, it is clear
that the enumeration becomes more and more complicated as $N$
increases.

\subsection{Slow detachment limit ($0< q \ll 1$)}

\noindent
Although mean assembly times are infinite in an irreversible process
(except when $M$ is odd and $N=3$), they are finite when $q>0$.  For
general values of $M$ and $N$ and for small $q>0$, we can find the
leading behavior of the mean first assembly time $T_N(M,0,\cdots, 0)$
perturbatively by considering the trajectories from nearly trapped states
into an absorbing state with at least one completed cluster.

Since the mean arrival time to an absorbing state is the sum of the
probabilities of each pathway, weighted by the time taken along each
of them, we expect that the dominant contribution to the mean assembly
time in the small $q$ limit can be approximated by the shortest mean
time to transition from a trapped state to an absorbing state. This
assumption is based on the fact that the largest contribution to the
mean assembly time will arise from the waiting time to exit a trap, of
the order of $ \sim 1 /q$, since detachment is the only possible path
out of the otherwise trapped state.  The time to exit any other state
will be of order 1 since monomer attachment is allowed. For
sufficiently small detachment rates $q$, we thus expect that the
dominant contribution to the mean assembly time comes from the trajectories
that sample nearly trapped states and that $T_N(M,0, \dots, 0) \sim 1/q$.

Again, first consider the tractable case $N=3$ and $M$ even, where it
is clear that the sole trapped state is $(0,M/2,0)$ and the
``nearest'' absorbing state is $(1,M/2-2,1)$. Since the largest
contribution to the first assembly time occurs along the path out of
the trap and into the absorbed state, we posit

\begin{equation*}
T_3(M,0,0) \simeq P^*(0, \frac M 2,0) \, T_3(0, \frac M 2,0),
\end{equation*}

\noindent
where $P^*(0,M/2,0)$ is the probability of populating the trap,
starting from the $(M,0,0)$ initial configuration for $q=0$.  This
quantity can be evaluated by considering the different weights of each
path leading to the trapped state. An explicit recursion formula has
been derived in our previous work \cite{DLC} in Section 4 and in
Eq.\,A.12. In the $N=3$ case however, the paths are simple, since only
dimers or trimers are formed, leading to

\begin{eqnarray}
\label{factor}
P^{*}(0, \frac M 2 ,0) = \frac{(M-3)!!}{(M-1)^{\frac M 2 - 1} M},
\end{eqnarray}

\noindent
which is the same as what was derived in Eq.\,\ref{W2}.  The first
assembly time $T(0,M/2,0)$ starting from state $(0,M/2,0)$ is

\begin{equation}
\label{FPT1}
\displaystyle{T_3(0,\frac M 2,0) = \frac 1 {\frac {M} {2} q} +
T_3(2,\frac M 2 -1, 0)}.
\end{equation}

\noindent
Here, the first term is the total exit time from the trap, given by
the inverse of the detachment rate $q$ multiplied by the number of
dimers. The second term is the first assembly time of the nearest and
sole state accessible to the trap.  This quantity can be evaluated, to
leading order in $1/q$, as

\begin{eqnarray}
\label{FPT2}
T_3(2,\frac M 2 - 1,0) \simeq
\frac 1 {2 (\frac M 2 - 1) + 1}  \,T_3(0, \frac M 2,0),
\end{eqnarray}

\noindent
where we consider that the trap will be revisited upon exiting the
state $(2, M/2-1,0)$ with probability $1 / (2 (\frac M 2 - 1) +
1)$. In principle, Eq.\,\ref{FPT2} should also contain another term
representing the possibility of reaching state $(4, M/2 -2, 0)$ via
detachment from state $(2,M/2-1,0)$ and its contribution to the first
assembly time. However, the magnitude of this term would be much
smaller than $1/q$, since detachment rates are of order ${\cal O}(q)
\ll {\cal O}(1/q)$.  Another term that should be included in
Eq.\,\ref{FPT2} is the possibility of reaching the absorbing state
$(1, M/2-2,1)$. This term however, yields a zero contribution to the
first assembly time.  Upon combining Eqs.\,\ref{FPT1} and \ref{FPT2}
we find that as $q \to 0$

\begin{equation}
%\label{FPT3}
\displaystyle{T_3(0,\frac M 2,0) \simeq
\frac {2 (M-1)} {M (M-2)} \frac 1 q}. \nonumber
\end{equation}

\noindent
Finally, $T_3(M,0,0)$ can be derived by multiplying the above result
by Eq.\,\ref{factor}.  We can generalize this procedure to find the
dominant term for the mean assembly time starting from \textit{any}
initial state $(M-2n,n,0)$ in the limit of small $q$, $N=3$ and for
$M$ even

\begin{widetext}
\begin{eqnarray}
\label{T3approx}
T_3(M,0,0) \simeq  T_3(M-2,1,0)& \simeq & 
\frac{2(M-3)!!} {M (M-2)(M-1)^{M/2-2}} \frac 1 q, \\
T_3(M-2n,n,0) & \simeq & \frac{2 (M-2n-1)!!} 
{M (M-2)(M-1)^{M/2-n-1}} \frac 1 q \quad  2\leq n < M/2  \label{Tb}\\
T_3(0,M/2,0) & \simeq & \frac {2(M-1)}{M(M-2)}\frac {1}{q}.
\end{eqnarray}
\end{widetext}

\noindent
The next correction terms do not have an obvious closed-form
expression, but are independent of $q$.  Note that when $q$ is small
and increasing, the mean first assembly times {\it decrease}. This is
also true for odd $M$. A larger $q$ leads to a more rapid
dissociation, which may lead one to expect a {\it longer} assembly
time. However, due to the multiple pathways to cluster completion in
our problem, increasing $q$ actually allows for more mixing among
them, so that at times, upon detachment, one can ``return'' to more
favorable paths, where the first assembly time is actually shorter.
This effect is clearly understood by considering the case of $q=0$
when, due to the presence of traps, the first assembly time is
infinite. We have already shown that upon raising the detachment rate
$q$ to a non-zero value, the first assembly time becomes finite. Here,
detachment allows for visiting paths that lead to absorbed states,
which would otherwise not be accessible. This same phenomenon persists
for small enough $q$ and for all $M,N$ values.  The expectation of
assembly times increasing with $q$ is confirmed for large $q$ values,
as we shall see in the next section.  Taken together, these trends
indicate the presence of a minimum in the mean first assembly time
that occurs at an intermediate value of the detachment rate $q$.

We can generalize our estimate of the leading $1/q$ term for the
first assembly time to larger values of $N$ via

\begin{eqnarray}
\label{general}
T_N(M,0, \dots, 0) = \sum_{\{\mu\}} P^*(\{\mu\}) T_{N}(\{\mu\}), 
\end{eqnarray}

\noindent
where $\{\mu\}$ 
are trapped state configurations for $q=0$.  The values of $P^*(\{\mu\})$ can
be calculated as described above using the recursion formula presented
in \cite{DLC}.  Approximate mean first assembly times $T_{N}(\{\mu\})$
from traps $\{\mu\}$ may be found by considering equations for the shortest
sub-paths that link traps to each other. For instance, in the case of
$M=9$, $N=4$ the only trapped states are $(0,3,1,0)$ and $(0,0,3,0)$,
with associated probabilities $P^*(0,0,3,0) = 921/5488$ and
$P^*(0,3,1,0) = 2873/24696$, respectively.  The shortest path linking
the two traps is $(0,3,1,0) \rightarrow (2,2,1,0) \rightarrow
(1,1,2,0) \rightarrow (0,0,3,0)$, which yields, to first order,
$T(0,1,3,0) = T(0,0,3,0) = 1 / (2q)$.  Finally, from
Eq.\,\ref{general} we find that $T(9,0,0,0) = 2005/ (14112 q)$ which
can be verified by constructing the corresponding $D(9,4) = 12$
dimensional transition matrix ${\bf A}^\dagger$ and solving the linear
eigenvalue problem. Enumerating trajectories that intersect nearly
trapped states becomes increasingly complex as $M$ and $N$ increase
since more traps arise, leading to the identification of more
entangled sub-paths connecting them.

\subsection{Fast detachment limit ($q \to \infty$)}

\noindent
We now consider the case where detachment is much faster than
attachment and $q \gg M$. In this limit, we expect the full assembly
of a cluster to be a rare event in the large $q$ limit, and that the
mean assembly time will increase monotonically with $q$. 

\vspace{3mm}

\noindent {\it Dominant path approximation - }Our first approximation
is based on the observation that for $q \to \infty$ the dominant
configurations are those with the most monomers (the higher states in
each column of Fig.~\ref{TREE}).  Thus, the dominant trajectories will
be the ones that most directly arrive at the absorbing state with one
full cluster. For $N=3$, the overwhelmingly dominant paths are:
$(M,0,0) \rightleftharpoons (M-2,1,0) \rightleftharpoons
(M-3,0,1)$. The dynamics of the probabilities of the two ``surviving''
states with $n_3 = 0$ can be represented by a linear $2 \times 2$
system that is easily solved to yield, in the $q \to \infty$ limit,

\begin{eqnarray}
%\label{eq:asymptot2}
T_3(M,0,0) &\simeq& T_3(M-2,1,0) \simeq \frac{2q}{M(M-1)(M-2)},  \nonumber
%\\ T_3(M-2,1,0) &\simeq& \frac{2q} {M(M-1)(M-2)},
\end{eqnarray} 

\noindent
The dominant path method can be generalized to any $M \geq N$ for $q \gg  M$
as follows

\begin{widetext}
\begin{equation}
\label{1cluster}
\begin{array}{cccccccccc}
(M,0,0,...,0) & \rightleftharpoons & (M-2,1,0...,0) &
\rightleftharpoons & \cdots & \rightleftharpoons (M-r,0...,1,..,0) &
\rightleftharpoons & \cdots & \rightleftharpoons &
(M-N,0,...0,1).
\end{array}
\end{equation}
\end{widetext}

\noindent
Here, the corresponding transition matrix ${\bf R}^{\dagger}$ is
tridiagonal and of dimension $(N-1)$ with elements $r^{\dagger}_{1,1}
= - r^{\dagger}_{1,2} = - M (M-1)/2$ and $r^{\dagger}_{k,k-1} = q,
r^{\dagger}_{k,k} = - q - (M-k), r^{\dagger}_{k,k+1} = (M-k)$ for $2
\leq k \leq (N-1) $.  The inverse of ${\textbf R}^{\dagger}$ can be
computed by a three-term recurrence formula \cite{USMANI}.  After some
algebraic manipulation, we can write the first assembly time along
the path in Eq.\,\ref{1cluster} for any $M\geq N$ and for $q \geq M$
as

\begin{widetext}
\begin{eqnarray}
 \label{tlag}
T_{N}(M,0, \dots, 0)
=\frac{2 q^{N-2}}{\prod_{i=0}^{N-1}(M-i)}&\Bigg{[}
&\sum_{k=0}^{N-2}\prod_{\ell=1}^{k}(M-(N-\ell))q^{-k}
\\ \nonumber &+&\frac{M(M-1)}{2}
\sum_{j=2}^{N-2}\prod_{\ell=2}^{j-1}
(M-\ell)\sum_{k=0}^{N-j-1}\prod_{l=1}^{k}(M-(N-\ell))q^{1-j-k}\Bigg{]}.
\end{eqnarray}
\end{widetext}

\noindent
Our notation is such that products with the lower index larger
than the upper one are set to unity.  In Eq.\,\ref{tlag}, the largest
term in the $q \to \infty$ limit is given by

\begin{equation}
%\label{eq:asymptot3}
T_{N}(M,0,\dots,0) \simeq \frac{2q^{N-2}}{\prod_{i=0}^{N-1}(M-i)}. \nonumber
\end{equation} 

\noindent
The additional assumption $M \gg N$ on the other hand, leads to the
 approximation $M-i\simeq M$ so that Eq.\,\ref{tlag} becomes

\begin{equation}
\label{approxMN}
T_{N}(M,0,\dots,0) \simeq \frac{q^{N-1}}{M^N}\Bigg{[}
  \sum_{k=2}^{N-1} \frac{(k-1)M^k}{q^k}
  +\frac {2}{q} \sum_{k=0}^{N-2}\frac{M^k}{q^k}\Bigg{]}.
\end{equation}

\noindent
Results for other choices of initial configurations $\{m\}$ can be
obtained by following the same reasoning illustrated here. We expect
$T_N(\{m\})$ not to be too different from $T_N(M,0,\dots,0)$ in the
strong detachment $q \to \infty$ case when any initial clusters will
rapidly disassemble, leading the system towards the free monomer
configuration. The distribution of first assembly times can also be
obtained within the dominant path approximation, as outlined in
Appendix \ref{APPENDIXC}.

We expect these results to hold for large $q \geq M$, small values of
$N$ and moderate values of $M$ so that the most likely trajectories
follow the dominant path. However, due the possibility of many
branching paths in configuration space, modest changes in $\{M,N,q\}$
may allow sampling of secondary paths that yield different estimates
of the first assembly time. Indeed, as both $M$ and $N$ become larger,
the creation of several intermediate clusters may be more favorable
than progressively adding monomers to the largest one.  In the next
subsection we thus introduce a ``hybrid'' approach, where the
possibility of having multiple intermediate aggregates is included by
assuming that the first $r$ clusters are distributed according to the
Becker-D\"oring equilibrium distribution and the remaining $N-r$
follow a monomer-to-largest cluster path towards complete assembly.

\vspace{3mm}

\noindent {\it Hybrid approximation - }We now consider a different
approach to the fast detachment $q \to \infty$ limit by using a
``pre-equilibrium'' or ``quasi steady-state'' approximation
\cite{powers06} that partially neglects correlations between some of
the cluster numbers by separating time scales between fast and slow
varying quantities.  We will use the pre-equilibrium approximation on
the stochastic formulation of Eq.\,\ref{MASTER}; however, to
illustrate the method, we will first apply it to
Eqs.\,\ref{HOMOEQN0}, with $q_N=0$

The equilibrium values $c^{\rm eq}_i$ arising from Eqs.~\ref{HOMOEQN0}
where $p_i=1$ are given by

\begin{equation}
%\label{eq:equilibre_determinist}
 c_i^{\rm eq}=\frac{(c_1^{\rm eq})^i}{2 q^{i-1}}, \nonumber
\end{equation}

\noindent
for $2 \leq i \leq N$, whereas $c_1^{\rm eq}$ can be obtained using the
mass conservation constraint

\begin{equation}
%\label{eq:massprop}
 c_1^{\rm eq}+\frac{1}{2}\sum_{i=2}^n \frac i 
{q^{i-1}}(c_1^{\rm
 eq})^i = c_1(0) = M. \nonumber
\end{equation}

\noindent 
We can now define the fluxes $J_i^{\pm}(t)$ 

\begin{eqnarray}
 J_i^+(t) & \equiv  c_1(t) c_{i-1}(t) &\simeq
 \frac{(c_1^{\rm eq})^i}{2 q^{i-2}},
 \nonumber\\ J_i^-(t) & \equiv qc_{i+1}(t) &\simeq
 \frac{ (c_1^{\rm eq})^i}{2 q^{i-2}}, \nonumber
\end{eqnarray}

\noindent
for $1 \leq i \leq N-1$. Note that as $q \to \infty$ all fluxes
decrease and that since $q$ is large, $J_i^{\pm}(t) \gg
J_{i+1}^{\pm}(t)$.  This implies that smaller clusters experience
faster dynamics and motivates the quasi-steady state approximation.  
We may thus consider the first $N-1$ reactions to be at
equilibrium so that Eqs.~\ref{HOMOEQN0} can be rewritten as a function
of the mass $x(t)$ contained in all clusters except the largest one:
 
\begin{equation}
 x(t)=\sum_{i=1}^{N-1} ic_i(t).
\end{equation}

\noindent
Eqs.\,\ref{HOMOEQN0} now become

\begin{eqnarray}
\dot x(t) &=& - N \dot c_N(t) = 
-\frac{ N}{2 q^{N-2}}c_1(x)^{N}, 
\label{eq:cn_det_reduce}
\end{eqnarray}

\noindent
where $c_1(x(t))$ satisfies

\begin{equation}\label{eq:preeq_n}
 c_1+\frac{1}{2}\sum_{i=2}^{N-1} \frac i {q^{i-1}}(c_1)^i=x.
\end{equation}

\noindent
Upon solving Eq.\,\ref{eq:preeq_n}, we can obtain $c_1(x)$, which can
then be inserted in Eq.\,\ref{eq:cn_det_reduce} to determine $c_N(t)$
and $x(t)$. Upon using the crude approximation $c_1(x)\approx x$ in
Eq.\,\ref{eq:cn_det_reduce} we can solve for $x(t)$ and find the time
at which $c_N=1$, or equivalently $x=M-N c_N =M-N$

\begin{eqnarray*}
\hspace{-0.2cm}
T_N(M,0,\ldots,0) = \frac{q^{N-2}}{N} \left[
\frac{N-1}{(M-N)^{N-1}} - \frac{N-1}{M^{N-1}}
\right].
\end{eqnarray*}

\noindent
A more accurate result can be found by 
allowing the sum in Eq.\,\ref{eq:preeq_n} to go to infinity 
so that

\begin{equation}
%\label{eq:monomeratequil}
 c_1(x) =\frac{-q+\sqrt{4 q x + q ^2}}{2}.\nonumber
\end{equation}

\noindent
Using this expression in Eq.~\ref{eq:cn_det_reduce}, the mean mass in
the fast clusters $x(t)$ can be explicitly computed.  Note that we
could as well have assumed that only the first $1 \leq r < N-1$
clusters equilibrate among themselves, and solved a reduced system of
$N-r+1$ equations.  We do not pursue this explicitly here, and move
directly to the stochastic version of pre-equilibration.

A separation of time scales can also be performed in stochastic
systems, where the basic assumptions for pre-equilibration are the
same as for the deterministic case. In particular, we require the
``fast'' subsystem to be ergodic and to possess a unique equilibrium
distribution.  The dynamics of the ``slow'' subsystem is then obtained
by averaging the fast variables over their equilibrium distribution;
the basic assumption is that while slow variables evolve, the fast
ones equilibrate instantaneously to their average values \cite{Kang}.
As we shall see, due to the equilibrium hypothesis, summing
Eq.\,\ref{MASTER} over the variables that constitute the fast
subsystem, will lead to the vanishing of all terms that do not modify
the slow variable, and all remaining terms will involve averages of
the fast variable \cite{Haseltine2005}.

Just as in the deterministic case, we allow the first $N-1$ cluster
sizes to equilibrate amongst each other and write the probability
distribution function using a mean-field approach

\begin{equation}
\label{anz}
 P(\{n\};t\vert \{m\}, 0)=P_{\rm eq}(\{n'\} \vert n_N) P(n_N;t\vert
 \{m\}, 0).
\end{equation}

\noindent For fixed $n_{N}$, $P_{\rm eq}(\{n' \} | n_N)$ represents the
equilibrium distribution function for the first, fast $\{n'\} = \{n_1,
\dots, n_{N-1}\}$ cluster sizes and

\begin{equation}
\label{Integr}
 P(n_N;t\vert \{m\}, 0)=\sum_{\{n'\}} P(\{n'\},n_N; t\vert \{m\}, 0)\, 
\end{equation}

\noindent
is the probability distribution for the last, slow cluster size $n_N$.
The sum in Eq.\,\ref{Integr} is to be performed over all values of
$\{n'\}$ such that mass conservation $\sum_i^{N-1} i n_i = M - N n_N$
is obeyed.  Note that while $P_{\rm eq}(\{ n'\} | n_N)$ does not
depend on the initial conditions of the $\{ n'\}$ clusters, it does
depend on $n_N$.
%
%since we are considering the equilibration of a
%system of $N-1$ clusters of total mass $M - N n_N$.  
%
%At this point thus, the separation between fast and slow variables is
%not quite complete, an issue we will tackle below.  
%
Upon inserting the ansatz in Eq.\,\ref{anz} in
Eq.\,\ref{MASTER} and performing the summation over all
configurations $\{ n'\}$ with fixed $n_N$, we find

\begin{widetext}
\begin{eqnarray}
\nonumber
\dot{P} (n_N;t\vert \{m\}, 0) &= & 
-\big{(} \langle n_1 n_{N-1} | n_N \rangle_{\rm eq} +q n_N
\big{)}P(n_N;t\vert \{m\}, 0) \\
&&
\label{MASTER_reduce}
 \\
\nonumber
&& +  \langle n_1 n_{N-1} | n_N - 1 \rangle_{\rm eq}
P(n_N-1;t\vert \{m\}, 0)+q(n_N+1)P(n_N+1;t\vert \{m\}, 0).
\end{eqnarray}
\end{widetext}

\noindent
In Eq.~\ref{MASTER_reduce}, we have used the notation

\begin{equation}
 \langle n_1 n_{N-1} | n_N \rangle_{\rm eq} = \sum_{\{ n'\}} n_1 n_{N-1}
 P_{\rm eq}(\{n'\} \vert n_N)
\end{equation}

\noindent
representing the equilibrium second moment $\langle n_{1} n_{N-1}\vert
n_{N}\rangle_{\rm eq}$, which is an average over all fast variables
with the added constraint that they have total mass $M - N n_N$.
Eq.\,\ref{MASTER_reduce} implies that $n_N$ follows a Markovian birth
and death process with birth rate $ \langle n_1 n_{N-1} | n_N
\rangle_{\rm eq}$ and a death rate $q n_N$. Starting at $n_N=0$ at
time $t=0$, the first birth event coincides with the first assembly
time so that the survival probability can be written as

\begin{equation}
\label{define}
S_{N-1}(t)= \exp \Big{[} -\int_0^t \langle n_1 n_{N-1} | n_N = 0 \rangle_{\rm eq}
   {\rm d}t \Big{]} \equiv e^{- \lambda t},
\end{equation}

\noindent where the ``$N-1$'' indicates that all clusters of size
$N-1$ and smaller have been pre-equilibrated. Having defined $\lambda
= \langle n_1 n_{N-1} | n_N = 0 \rangle_{\rm eq}$ in
Eq.\,\ref{define}, the first assembly time distribution is exponential

\begin{eqnarray}
\label{esp}
G_{N-1}(\{m\};t) = \lambda e^{-\lambda t},
\end{eqnarray}

\noindent
and the mean first assembly time is given by $T_N(M,\cdots, 0) =
1/\lambda$.  The remaining difficulty lays in determining the quantity
$ \langle n_1 n_{N-1} | n_N \rangle_{\rm eq}$. We may resort to a very
crude approximation, by simply using the Becker-D\"oring results

\begin{eqnarray}
\label{BDapp}
 \langle n_1 n_{N-1} | n_N \rangle_{\rm eq} & \simeq & 
c_1^{\rm eq} c_{N-1}^{\rm eq} \\
%\langle n_1 \rangle \langle
%n_{N-1} \rangle \\
%& \approx & \frac{1}{2}\Big{(}\frac{p}{q}\Big{)}^{N-2} \langle
%n_1 \rangle ^N \nonumber \\
& \simeq & \frac{1}{2 q^{N-2}}(c_1^{\rm eq})^{N}.
\nonumber
\end{eqnarray}

\noindent
Eq.\,\ref{BDapp} can now be used to estimate $\lambda$ and all other
related quantities. Our work so far implies that the first assembly
time is exponentially distributed according to Eq.\,\ref{esp}.
However, upon comparing with results from Monte-Carlo simulations in
the next section, we will show that the $N-1$ pre-equilibration and is
often not a good approximation. As outlined in Appendix
\ref{APPENDIXD}, a less drastic approximation can be implemented by
allowing only the first $r$ species ($1 \leq r < N$) to
pre-equilibrate. This more restricted pre-equilibration approximation
can occasionally provide better fits to simulation as we will see in
the next section.

\section{Comparison with Simulations}

\noindent
In this section we present results derived from simulations of the
stochastic process associated to the probability distribution process
for various values of $\{M,N,q\}$.  Specifically, we use an exact
stochastic simulation algorithm (kinetic Monte-Carlo, KMC) to
calculate first assembly times \cite{BKL75, GILLESPIE77}. For each set
of $\{M,N,q\}$ we sample at least $10^4$ trajectories and follow the
time evolution of the cluster populations until $n_N=1$, when the
simulation is stopped and the first assembly time recorded.  We
compare and contrast our numerical results with the analytical
approximations evaluated in the previous sections.

\begin{figure}[t]
\centering
\includegraphics[width=3.4in]{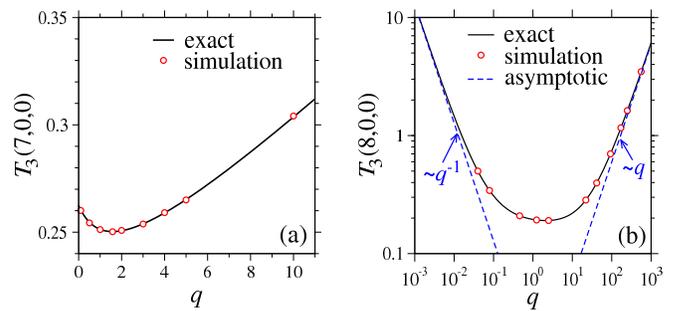}
\caption{Mean first assembly times for $M=7$ and $N=3$ in panel (a)
  and $M=8$ and $N=3$ in panel (b).  Exact results derived in
  Eq.\,\ref{analytic} are plotted as black solid lines, while red
  circles are obtained by averaging over $10^5$ KMC trajectories.  
The dashed blue
  line shows the $q \to 0$ approximation in Eq.\,\ref{T3approx}
  and the $q \to \infty$ approximation in Eq.\,\ref{tlag}.}
\label{M7}
\end{figure}

We begin with the simple case of $M=7$ and $N=3$ in Fig.\,\ref{M7}(a)
where we plot the mean first assembly time $T_3(7,0,0)$ as a function
of $q$ obtained via our exact results Eq.\,\ref{analytic} and by runs
of 10$^5$ KMC trajectories. Numerical and exact analytical results are
in very good agreement, in contrast to the discrepancies between the
fully stochastic and mean field treatments observed in
Fig.\,\ref{HEUR_N=3}.
%
%upon comparing the same exact analytical results to the estimates
%derived from the Becker-D\"oring system in Eqs.\,\ref{HOMOEQN0}.  We
%thus confirm the inadequacy of the heuristic method.
%
For comparison, we also plot in Fig.\,\ref{M7}(b) the mean
first assembly time $T_3(8,0,0)$ for $M=8$ and $N=3$, where the
presence of the trapped state $(0,4,0)$ leads to a diverging first
assembly time for $q=0$ and to the asymptotic behavior $T_3(8,0,0)
\sim 1/q$ for $q \to 0$, as predicted. Note that as discussed above
$T_3(7,0,0)$ is finite for $q=0$ due to the lack of trapped states for
$N=3$ and $M$ odd.  We do not plot the first assembly time
distributions as their features are similar to ones we will later
discuss.
%Here, we only note that 
%distributions for $M=7, N=3$ and
%$M=8, N=3$ are very similar.  The only noticeable difference is that for
%$M=7$ as $q \to 0$ the distribution is unimodally distributed around
%relatively short values of $t$, while an additional minor spread centered at
%$t \sim 1/q$ arises in the case of $M=8$.  This tail is due to runs
%that incur into the trapped state at $(0,4,0)$: although this is a
%rare event its occurrence increases the mean first assembly time
%significantly.  As $q$ increases, any differences between the
%``trap-free'' and ``trapped''distributions at $M=7$ and $M=8$ are
%blurred and both distributions approach a single parameter exponential
%with decay rate $\lambda_3$ as estimated by Eq.\,\ref{eq65}.

\begin{figure}[t]
\centering
\includegraphics[width=3.15in]{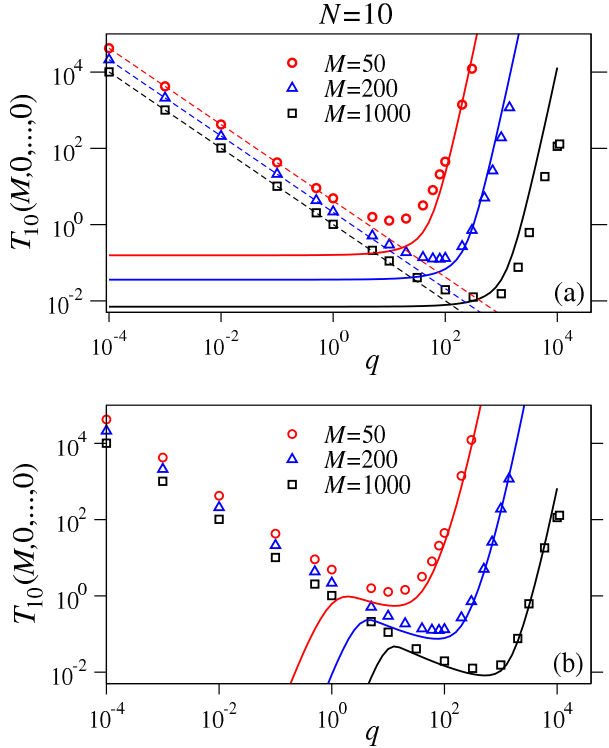}
\caption{Comparison of theory with simulations for $N=10$, and several
  values of $M$. Symbols are derived from $10^4$ KMC simulations for
  $M=50,200,1000$. In panel (a) the dashed lines are obtained by
  plotting the curve $T_{10}(M,0,\dots,0) = A/q$ where $A$ is given by
  imposing passage through the first point to the left in the
  graph. Note that all other points align to the same curve. Solid
  lines are derived from Eq.\,\ref{tlag} in the dominant path
  approximation. In panel (b) results from the hybrid approximation
  with $r=2$ in Eq.\,\ref{start} are superimposed on the same
  data. Note the much better fit in the hybrid approximation as $q \to
  \infty$, especially as $M$ becomes larger.}
\label{N10}
\end{figure}

We generalize this analysis by plotting numerical estimates of
$T_{10}(M,0, \cdots, 0)$ as a function of $q$ for various values of
$M$ in Fig.\,\ref{N10}(a). As expected, for small $q$, the mean first
assembly time scales as $1/q$ for all values of $M$. Similarly, for
all values of $M$, the first assembly time presents a minimum, due to
the previously-described increased weighting of faster pathways upon
increasing $q$ for small enough values of $q$. For larger values of
$q$ we expect the most relevant pathways towards assembly to be the
ones constructed along the linear chain described in
(\ref{1cluster}). Indeed, we find that in accordance with
Eq.\,\ref{approxMN}$, T_{N}(M,0,\cdots, 0) \simeq 2 q^{N-2} / M^N$ as
$q \geq M$.  Small and large $q$ estimates using the dominant path
approximation are shown in Fig.\,\ref{N10}(a).

As discussed earlier, the dominant path approximation becomes less
accurate as $M$ increases, since the linear chain pathway neglects
other possible routes towards complete assembly, that become relevant
as $M$ increases. In Fig.\,\ref{N10}(b) thus we plot the same data
points, using the hybrid approximation discussed above for large $q$,
with $r=2$. Note a much closer fit with the simulation data,
especially as $M$ increases.

\begin{figure}[t!]
\centering
\includegraphics[width=2.95in]{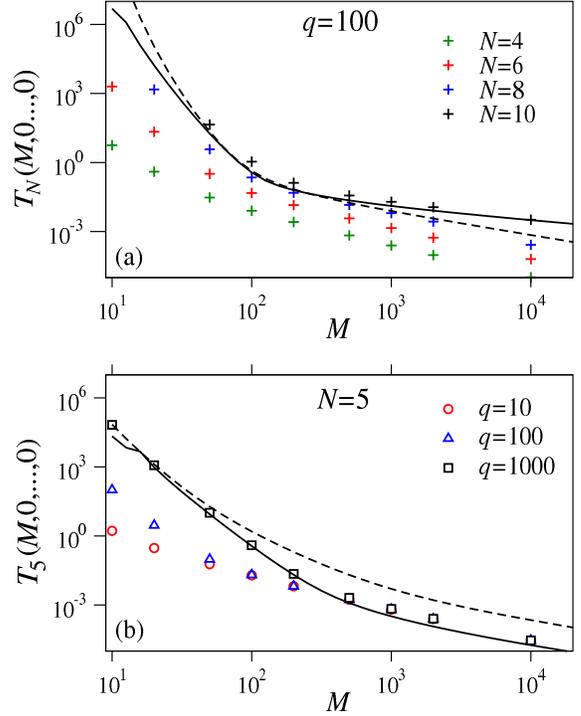}
\caption{First assembly times $T_N(M,0,\dots,0)$ as a function of $M$
  for $q=100$ and several values of $N$ in panel (a), and for $N=5$
  and several values of $q$ in panel (b). The black dashed lines
  represent the dominant path approximation for large $q$ in
  Eq.\,\ref{tlag}, while the solid black line represents the hybrid
  approximation in Eq.\,\ref{start} for $r=2$.  We chose to plot only
  representative cases, not to clutter the graphics, but similar
  trends persist in panel (a) for $N=4,6,8$ and in panel (b) for
  $q=10,100$.  Note that the dominant path approximation ceases to be
  accurate for very large values of $M$ and that the hybrid
  approximation provides a better fit as $q \to \infty$.}
\label{M8}
\end{figure}

In Fig.\,\ref{M8}(a) we plot $T_N(M,0,\cdots, 0)$ as a function of $M$
for $q$ fixed and various $N$, while in Fig.\,\ref{M8}(b)
$T_N(M,0,\dots,0)$ is plotted as a function of $M$ for $N$ fixed and
various $q$. Both Figs.\,\ref{M8}(a) and \ref{M8}(b) show that the
results derived in Eq.\,\ref{tlag} for large $q$ using the dominant
path approximation are accurate provided $M$ is not too large compared
to $N$. As shown by the black solid lines, in this case
$T_N(M,0,\cdots,0) \simeq 2 q^{N-2}/ M^N$.  For larger values of $M$,
the dominant path approximation becomes inaccurate: numerical results
indicate that $T_N(M,0,\cdots,0) \simeq 1/M^{\nu}$ with $\nu \sim 1$
as $q \to \infty$.  In this regime, the hybrid approximation with
$r=3$ yields a better fit, as shown by the solid lines in
Figs.\,\ref{M8}(a) and \,\ref{M8}(b).

%FINISH - FIND SOME ANALYTICAL JUSTIFICATION FOR THIS

\begin{figure}[t!] 
\centering
\includegraphics[angle=0,width=\linewidth]{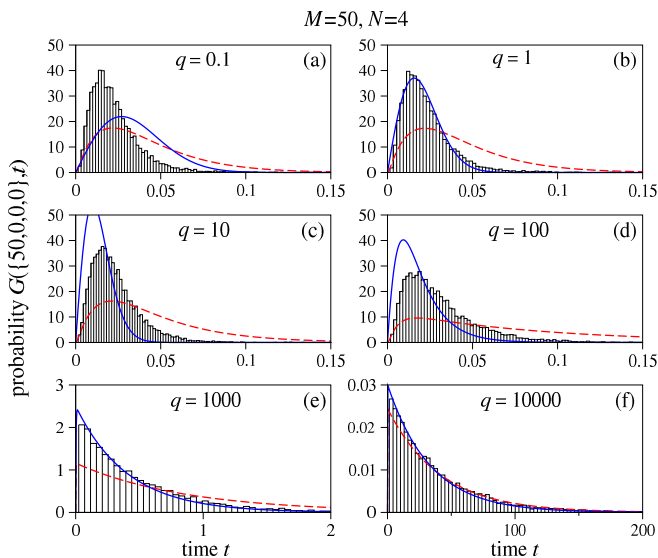}
\caption{Probability distributions for the first assembly time for
  $N=4$ and $M=50$ and for various values of $q$.  The black bars are
  obtained as a normalized histogram of $10^4$ KMC simulations. The
  dashed red and solid blue lines are the probability density
  functions estimated via the dominant path approximation in
  Eq.\,\ref{eq:laplace} and via the hybrid approximation with $r=3$ in
  Eq.\,\ref{start}, respectively.  The detachment rate $q$ increases
  as indicated in each subplot. Note that initially the distribution
  has a log-normal shape and later turns into an exponential. As
  predicted, the analytical estimate given by Eq.\,\ref{eq:laplace}
  becomes accurate for $q \geq M$.  Also note the change in scale and
  the broadening of the distribution as $q$ increases.}
\label{fig:5}
\end{figure}

\begin{figure}[t!]
\centering
\includegraphics[angle=0,width=\linewidth]{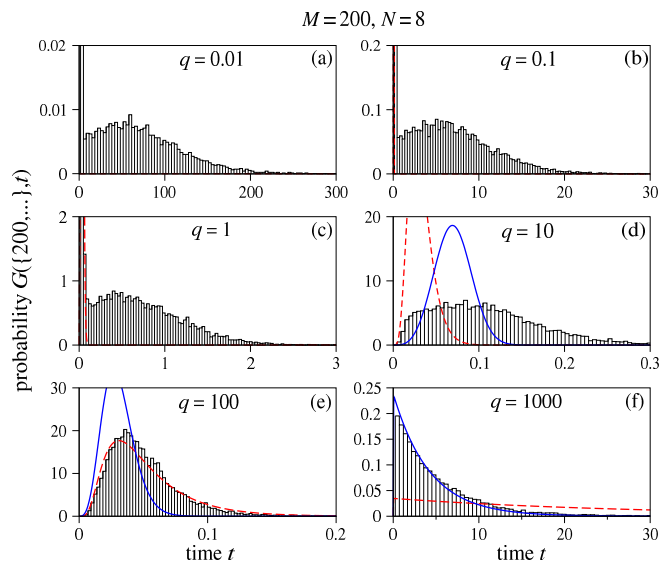}
\caption{First assembly time distributions for
  $N=8$ and $M=200$ for various values of $q$.  The black bars are
  obtained as a normalized histogram of $10^4$ KMC simulations. The
  dashed red and solid blue lines are the probability density
  functions estimated via the dominant path approximation in
  Eq.\,\ref{eq:laplace} and via the hybrid approximation with $r=3$ in
  Eq.\,\ref{start} respectively.  The detachment rate $q$ increases as
  indicated in each subplot. Note that the distribution evolves from a
  bi-stable curve to, later acquiring a log-normal shape, 
  before turning to an exponential.  As above, as $q \to \infty$ the hybrid
  approximation yields a closer fit to the numerical data.}
\label{fig:6}
\end{figure}

\begin{figure}[t!]
\centering
\includegraphics[angle=0,width=\linewidth]{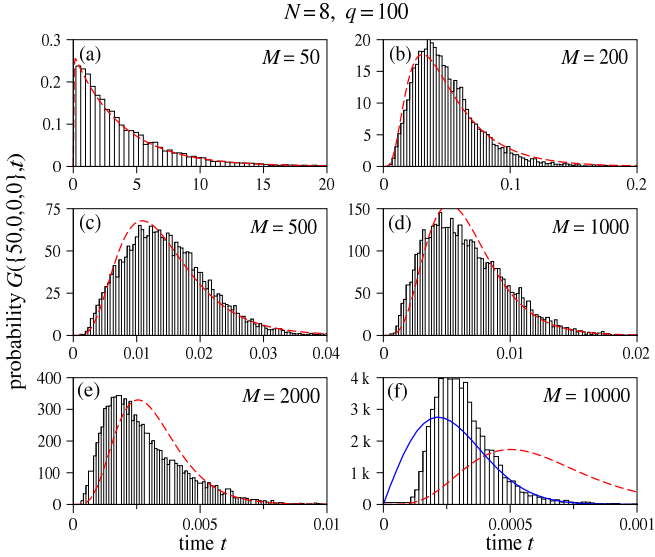}
\caption{First assembly time distributions for $N=8$ and $q=100$ for
  various values of $M$.  The black bars are obtained as a normalized
  histogram of $10^5$ KMC simulations. The dashed red and solid blue
  lines are the probability density functions estimated via the
  dominant path approximation in Eq.\, (\ref{eq:laplace}) and via the
  hybrid approximation with $r=3$ in Eq.\,\ref{start} respectively.
  Total mass $M$ increases as indicated in each subplot. Note that the
  distribution evolves from an exponential with decay rate given by
  Eq.\,\ref{eq65}, valid for $q \geq M$, towards a more log-normal
  shape.  In this case, for very large $M$ both dominant path and
  hybrid approximation fail.}
\label{fig:7}
\end{figure}

Finally, in Figs.\,\ref{fig:5}--\ref{fig:7} we plot the 
distribution function $G(\{M,0,\cdots,0\},t)$ of the first
assembly times for several representative choices of $\{M,N,q\}$.  As
illustrated in the figure captions, analytical estimates were
calculated either by inverse Laplace transforming
Eq.\,\ref{eq:laplace} after having numerically found its poles, or via
the hybrid approximation in Eq.\,\ref{start} with specific values of
$r$.  From Fig.\,\ref{fig:5} note upon
increasing $q$, $G(\{M,0,\dots,0 \},t)$ gradually shifts from having a
log-normal shape towards an exponential distribution characterized by
the decay rate evaluated in Eq.\,\ref{eq65}.  Some combinations of $M$
and $N$, such as $M=200$ and $N=8$ in Fig.\,\ref{fig:6} yield a
bimodal distribution for small $q$. This can be explained by noting
that while fast routes towards nucleation may exist, other pathways
lead the system to the previously described trapped states where
$n_1=n_N=0$.  Exit from these traps is unlikely for small $q$,
yielding larger first assembly times. The emergence of a bimodal
distribution should be more apparent for larger values of $N$ when
there is a longer pathway towards assembly and more potential traps.
Indeed, although not shown in Fig.\,\ref{fig:5} for $M=50$ and $N=4$,
a few trajectories populate the region $t \sim 1/q$, indicating passage
through at least one of the nine possible trapped states.  However,
the weights of these possible paths are very small (only about 10 or so
out of $10^{4}$ runs incurred into a trapped state), so we do not
include them in Fig.\,\ref{fig:5} which is truncated at $t \ll 1/q$,
when $q \to 0$. This occurs also for $M=8$ and $N=3$, where a minor
spread due to the $(0,4,0)$ trap and centered around $t \sim 1/q$
arises in the distribution tail, and which is absent from the
trap-free case of $M=7$ and $N=3$.

Note that although few paths may populate the region $t \sim 1/q$ their
contribution to the mean first assembly time may be significant.  In
Fig.\,\ref{fig:6} we also include analytical estimates of the first
assembly times: the dashed red curves are derived from the dominant
path approximation in Eq.\,\ref{eq:laplace} and the solid blue ones
from the hybrid approximation in Eq.\,\ref{start} using $r=3$.  
As noted above,  for very large $q$, the dominant path
approximation fails and the hybrid approximation provides a
closer fit to our numerical results.

In Fig.\,\ref{fig:7}, we plot the first assembly time distribution for
fixed $q=100$ and $N=8$ and varying $M$. As expected, for $q \geq M$,
$G(\{M,0,\cdots,0 \},t)$ is well approximated by the exponential
distribution in Eq.\,\ref{eq66}.  As $M$ increases the distribution
acquires a log-normal shape.  In this case, we find the hybrid
approximation to fail regardless of $r$.  Indeed, our numerical
results show that there is no specific criterion to ensure that the
hybrid approximation will yield even qualitatively valid estimates for
the first assembly distributions as $M \to \infty$.  Empirically, we
find that while mean first assembly times predictions are quite
accurate within the hybrid approximation, the first assembly
distribution estimates are more likely to be accurate when the are
exponentially distributed.

%IS THE BIMODAL ALSO DEPENDENT ON M SOMEHOW?
%TRY M=200, N=4.

%T_{8,0,0} & = & \frac{105 + 1526 q + 488 q^{2} + 40 q^{3} + q^{4}}
%{168 q (49 + 16 q + q^{2})}\\ T_{6,1,0} & = & \frac{105 + 1232 q + 392
%q^{2} + 34 q^{3} + q^{4}} {168 q (49 + 16 q + q^{2})}\\ T_{4,2,0} & =
%& \frac{147 + 1176 q + 350 q^{2} + 30 q^{3} + q^{4}} {168 q (49 + 16 q
%+ q^{2})}\\ T_{2,3,0} & = & \frac{343 + 1386 q + 350 q^{2} + 28 q^{3}
%+ q^{4}} {168 q (49 + 16 q + q^{2})}\\ T_{0,4,0} & = & \frac{2401 +
%2058 q + 392 q^{2} + 28 q^{3} + q^{4}} {168 q (49 + 16 q + q^{2})}
%\end{eqnarray}

\section{Summary and Conclusions}

\noindent
We have studied the problem of determining the first assembly time of
a cluster of a pre-determined size $N$ to form from an initial pool of
$M$ independent monomers characterized by uniform attachment and
detachment rates $p = 1$ and $q$, respectively. We have shown that
while heuristic approaches using the traditional Becker-D\"oring
equations can be developed, these fail to capture relevant qualitative
features, such as divergences and non-monotonic behavior. A full
stochastic approach, based on the Backward Kolmogorov equation, 
was investigated.

We developed our stochastic model and were able to find exact results
for the first assembly time in systems where $M,N$ are small enough
for analytical treatments to be feasible.  For general $M,N$ we were
able to estimate general trends and behaviors for both large and small
$q$.  In particular, we find that in the absence of detachment, when
$q=0$, trapped states arise from which the system is never able to
escape, leading to infinitely large first assembly times. Furthermore,
we showed that these traps arise for all values of $N > 3$, regardless
of $M$. The possibility of a trap, and of diverging first assembly
times is not captured by the heuristic approach, and is confirmed by
our KMC simulations.  We are also able to show that for small $q$, the
divergence in the first assembly time scales as $1/q$. The latter
result may appear counter-intuitive, since larger detachment rates
should intuitively hinder the assembly process, leading to the
expectation that larger $q$ implies larger first assembly times.
While this is true in the $q \to \infty$ limit, in the case of $q \to
0$ an opposite trend arises: the increased accessibility of potential
paths in configuration space that lead to more rapid first assembly
times. As $q$ increases, these new paths become increasingly
populated, yielding an overall decrease in the first assembly time.
Finally, for larger values of $q$ we identify the most likely path to
be traveled in phase space towards the first assembly of an
$N$-cluster and derive estimates for the associated first assembly
time and probability distribution functions. For $q \gg 1$, we also
considered a ``hybrid'' approach where the first few clusters were
allowed to equilibrate, while the larger ones were still evolving
stochastically. In certain cases we were able to find better agreement
with numerical data, while for other combinations of $\{M,N,q \}$ the
hybrid approach fails.  The collection of analytic approaches for the
limits $q=0$, $0< q \ll 1$, and $q\gg 1$ are outlined in Sections
IV A, B, and C, respectively.

All of our analytical results were confirmed by our KMC simulations,
from which we obtained first assembly times and related probability
distribution functions. For certain choices of $\{M,N,q\}$ the
presence of traps could be indirectly inferred by the emergence of
bimodal distributions with very large first assembly times (on paths
where traps were encountered) and very short ones (on others that were
able to avoid them).  These bimodal distributions may be smeared out
for other choices of $\{M,N,q \}$.

A number of additional stochastic properties of our self-assembly
problem can be calculated.  For example, one can derive analogous
results for attachment and detachment rates $p_{k}$ and $q_{k}$ that
depend on cluster size $k$.  In particular, if we assume that binding
and unbinding of monomers depends on the available surface area, and
that clusters are of spherical shape, we can use the forms $p_k, q_k
\sim k^{2/3}$. Similarly, one could assume that stoichiometric
limitations could exist so that attachment of monomers becomes
progressively slower as completion of the $N$-mer is approached so
that $p_k \sim (N-k)$ and $q_k \sim k$.  These extensions as well as
the treatment of heterogeneous nucleation and first ``breakup'' times
will be considered in future work.

\section{Acknowledgements}

R.Y. was supported by ANR grant 08-JCJC-0135-01.  This work was also
supported by the National Science Foundation through grants
DMS-1021850 (MRD) and DMS-1021818 (TC).  MRD was also supported by an
ARO MURI grant (W1911NF-11-10332), while TC was also supported by ARO
grant 58386MA.

\begin{appendix}

\section{Calculation of Survival probability and moments}
\label{APPENDIXA}

To obtain expressions for moments of the first assembly
time, it is useful to Laplace transform Eq.\,\ref{gi} so that

\begin{equation}
%\label{giLapla}
\tilde {\bf G} = {\bf 1} - s \tilde {\bf S}. \nonumber
\end{equation}

\noindent
Here, $\tilde {\bf G}$ and $\tilde {\bf S}$ are Laplace transforms
of ${\bf G}$ and ${\bf S}$, respectively. The vector ${\bf 1}$ is the
survival probability of any initial, nonabsorbing state, and consists
of 1's in a column of length given by the dimension of ${\bf
  A}^{\dagger}$ on the subspace of nonabsorbing states. Using this
representation we may evaluate the mean assembly time for forming the
first cluster of size $N$ starting from the initial configuration
$\{m\}$

\begin{eqnarray}
\label{timeone}
T_{N}(\{m\}) & \equiv &  -\int_{0}^{\infty} t 
\frac{\partial S(\{m\};t)} {\partial t} {\rm d}t \\
& = & \int_{0}^{\infty} S(\{m\};t){\rm d}t \nonumber 
= \tilde{S}(\{m\};s=0),
\end{eqnarray}

\noindent
Similarly, the variance ${\rm V}_N(\{m\})$ 
of the first assembly time can be expressed as

\begin{eqnarray}
\nonumber
{\rm V}_N (\{ m \}) &\equiv & - \int_0^{\infty} t^2
\frac{\partial S(\{m\};t)}{\partial t} {\rm d} t - T_N^{2}(\{ m \}) \\
\nonumber
&=& 2 \int_0^{\infty} t S(\{ m\};t) {\rm d} t - T_N^{2}(\{ m \}) \\
\label{varone}
& =& -2 \,\frac{\partial \tilde S(\{m\},s)}{\partial s}\bigg|_{s=0} -
\tilde{S}^{2}(\{m\};0).
\end{eqnarray}

\noindent
Upon Laplace-transforming $\dot{\bf S} = {\bf A}^{\dagger}{\bf S}$ and
applying the initial condition $S(\{m\};t=0)= {\bf 1}$, we
find

\begin{eqnarray}
\label{AI0}
\tilde{\bf S} &=&  [s {\bf I} - {\bf A}^{\dagger}]^{-1}{\bf
  1},
\end{eqnarray}

\noindent
so that 

\begin{equation}
%\label{giLapla2}
\tilde {\bf G} = {\bf 1} - s [s {\bf I} - {\bf A}^{\dagger}]^{-1}{\bf
  1}. \nonumber
\end{equation}

\noindent
The first assembly time starting from a specific configuration
$\{m\}$ is thus 

\begin{eqnarray}
T_{N}(\{m\}) &=& \tilde{S}(\{ m\};0) = 
-\left[({\bf A}^{\dagger})^{-1}{\bf 1} \right]_{\{m\}}, 
\label{AI}
\end{eqnarray}
\noindent
where the subscript $\{m\}$ refers to the vector element
corresponding to the $\{ m \}^{\rm th}$ initial configuration.
Similar expressions can be found for the variance and other moments.

In order to invert the matrix ${\bf A}^{\dagger}$ on the subspace of
non-absorbing states we first note that its dimension $D(M,N)$ rapidly
increases with $M$. In particular, we find that the number of
distinguishable configurations with no maximal cluster obeys the
recursion

\begin{equation}
D(M,N+1) = \sum_{j=0}^{\left[M/N\right]}D(M-jN,N),
\label{RECURSIOND}
\end{equation}
where $\left[M/N\right]$ denotes the integer part of $M/N$.  For
example, in Eq.~\ref{RECURSIOND}, $D(M,2)=1$, and the only
``surviving'' configuration not to have reached at least one cluster
of size $k=2$ is $(M,0)$.  The next term is $D(M,3) = 1+
\left[M/2\right]$ which, for $M \to \infty$ yields $D(M,3) \simeq
M/2$. Similarly, $D(M,4)$ can be written as

\begin{eqnarray*}
%\hspace{-0.4cm} 
D(M,4) = \sum_{j=0}^{\left[M/3\right]} D(M-3j,3)
\simeq \left[ \frac M 3 \right] \left[\frac M 2 \right] \simeq
\frac {M^{2}} {6},
\end{eqnarray*}

\noindent
where the last two approximations are valid for large $M$.
By induction, we find

\begin{equation}
D(M,N) \simeq \frac {M^{N-2}} {(N-1)!}. \nonumber
\end{equation}

\noindent
From these estimates, it is clear that the complexity of the
eigenvalue problem in Eq.\,\ref{AI} increases dramatically for large
$M$. This enumeration of states and the associated matrix method for
computing first assembly times is analogous to the study of first
passage times on a network \cite{benichou}. However, rather than
considering statistical properties of a scale free network, we are
concerned with a probability flux across a specific realization of
a state space network.

We begin by studying the case of $N=3$ for general $M$ where
instructive explicit solutions can be derived for the mean assembly
times.  In this case, the eigenvalue problem for the vector of
survival probabilities ${\bf S} \equiv (S(M,0,0;t), S(M-2, 1,0;t),
S(M-4, 2, 0; t), \ldots)$ can be written using a tridiagonal
transition matrix ${\bf A}^{\dagger}$ whose elements
$a^{\dagger}_{i,j} = a_{j,i}$ take the form

\begin{eqnarray}
\nonumber
a^{\dagger}_{k,k-1} &=& (k-1)q_{2}, \quad 2 \leq k \leq 1 + \left[
\frac M 2 \right]
\end{eqnarray}

\begin{eqnarray}
\nonumber
a^{\dagger}_{k,k} &=& -\frac{(M-2k + 2)(M-2k+1)}{2} p_{1} - (k-1) q_2 - \\
\nonumber
&& (k-1) (M-2k+2)p_{2}, \quad \\
&& \hspace{2.7cm} 1\leq k \leq  1 + \left[ \frac M 2 \right] 
\nonumber
\end{eqnarray}

\begin{eqnarray}
\nonumber
a^{\dagger}_{k,k+1} &=& \frac {(M-2k+2)(M-2k+1)}{2}p_{1}, \quad \\
&& 
\nonumber
\hspace{2.7cm} 2\leq k \leq 1 + \left[ \frac M 2 \right]
\end{eqnarray}

\noindent 
where the first (second) index denotes the column (row) of the matrix.
Using the above form for ${\bf A}^{\dagger}$, we can now symbolically
or numerically solve for the Laplace-transformed survival probability
${\tilde S}(\{m\};s)$ and the mean self-assembly time ${\tilde
  S}(\{m\};s=0)$.

\section{Calculation procedure for irreversible limit $q=0$}
\label{APPENDIXB}

When $q=0$, the matrix ${\bf A}^{\dagger}$ now becomes bi-diagonal and
a two-term recursion can be used to solve for the survival probability
$\tilde S(M-2n,n,0; s)$ as follows. If the entries of the bidiagonal
matrix ${\bf A}^{\dagger}$ are denoted $a^{\dagger}_{ij}$, then the
elements $b_{i,j}$ of the inverse matrix ${\bf B} = \left[s {\bf I} -
  {\bf A}^{\dagger}\right]^{-1}$ are given by

\begin{eqnarray}
\nonumber
b_{i,i} &=& \frac 1 {s - a^{\dagger}_{i,i}},\\ 
b_{i,j} &=& 0 \quad \mbox{if $i > j$},\\ 
\nonumber
b_{i,j} & =& 
  \displaystyle{\frac{\prod_{k=i}^{j-1} a^{\dagger}_{k,k+1}} 
{\prod_{k=i}^{j} (s- a^{\dagger}_{k,k})}} \quad \mbox{if $i < j$}, 
\label{AIJ}. 
\\
\nonumber
\end{eqnarray}

\noindent
The Laplace-transformed survival probability, according to
Eq.\,\ref{AI0} is the sum of entries of each row of $\left[s {\bf I} -
  {\bf A}^{\dagger}\right]^{-1}$ 

\begin{equation}
\label{tofeed0}
\tilde S(M-2n, n, 0; s) = \frac{1}{s - a^{\dagger}_{i,i}} +\!\!
\sum_{j= i+1}^{[M/2]+1} \!\!\displaystyle{\frac{\prod_{k=i}^{j-1}
    a^{\dagger}_{k,k+1}}
{\prod_{k=i}^{j} (s - a^{\dagger}_{k,k})}},
\end{equation}

\noindent
\noindent
where $i = n+1$ is the $(n+1)^{\rm st}$ row of $\left[s {\bf I} - {\bf
    A}^{\dagger}\right]^{-1}$. Upon performing the inverse Laplace
transform of Eq.\,\ref{tofeed0} we can write the survival probability
$S(M-2n,n,0;t)$ as a sum of exponentials and derive the full first
assembly time distribution $- \partial S(M-2n,n,0;t)/ \partial t$.
Similarly, the mean first assembly time, according to Eq.~\ref{AI}, is
$T_3(M-2n,n,0) = \tilde S(M-2n,n,0; s=0)$.
%
%\begin{equation}
%\label{tofeed}
%T_3(M-2n, n, 0) = - \frac{1}{a^{\dagger}_{i,i}} +
%\sum_{j= i+1}^{[M/2]+1} (-1)^{i+j} \displaystyle{\frac{\prod_{k=i}^{j-1}
%    a^{\dagger}_{k,k+1}}
%{\prod_{k=i}^{j} a^{\dagger}_{k,k}}},
%\end{equation}
%\noindent
In particular, from Eq.~\ref{AIJ} we find

\begin{equation}
%\label{ratio}
\frac{a^{\dagger}_{k,k+1}}{a^{\dagger}_{k+1,k+1}}
= - \frac{(M-2k+2)(M-2k+1)}{(M-2k)(M-1)}, \nonumber
\end{equation}
which, when inserted into Eq.~\ref{tofeed0} with $s=0$ leads to
Eq.~\ref{TN3}.

\section{Calculation procedure for fast detachment $q\gg 1$}
\label{APPENDIXC}

An estimate for the first assembly time {\it distribution} can be
obtained within the dominant path assumption (Eq.~\ref{1cluster}).  By
using the symmetry properties of the associated matrix ${\textbf
  R}^{\dagger}$ we can find the Laplace transform of the first
assembly time distribution $G(\{M,0,\dots,0\};s)$ \cite{CHEMLA} in the
$q \geq M$ limit

\begin{equation}
\label{eq:laplace}
\tilde G(\{M,0,\dots,0\}; s)=\frac{\frac{1}{2}\prod_{i=0}^{N-1}(M-i)}{d_{N-1}(s)},
\end{equation}

\noindent
where $d_{N-1}(s)$ is a unitary polynomial 
of degree $N-1$, given by the following recurrence

\begin{eqnarray}
\nonumber
 d_1&=&s+\frac{M(M-1)}{2}, \\
\label{DISTR}
 d_2&=&(s+(M-2)+q)d_1 -q\frac{M(M-1)}{2}, \\
%=s^2+\big{(}\frac{M(M-1)}{2}+q+(M-2)
%\big{)}s+\frac{1}{2}\prod_{i=0}^{2}(M-i) \\
\nonumber
 d_{i} & = &(s+(M-i)+q)d_{i-1}-q(M-(i-1))d_{i-2}, \\
\nonumber
&& \mbox{    for $i >2$}.
%,\,\,\mbox{for \,\,\, i=3..N-1}
\end{eqnarray}

\noindent
Thus, $d_{N-1}(s)=s^{N-1}+...+\beta s^2 +\alpha s
+\frac{1}{2}\prod_{i=0}^{N-1}(M-i)$. 
Note that the first assembly time is given by

\begin{eqnarray*} 
T_N(M,0,\dots,0) = 
\lim_{s \to 0} \frac{1- \tilde G(\{M,0,\dots,0\};s)}s.
\end{eqnarray*}

\noindent
By comparing Eq.\,\ref{eq:laplace} with Eq.\,\ref{tlag} we note that
the term $\alpha$ that appears in the above expansion for
$d_{N-1}(s)$, corresponds to the quantity in the square brackets in
Eq.\,\ref{tlag} so that

\begin{eqnarray*}
T_{N}(M,0,\dots,0) = \frac{2 \alpha}{ \prod_{i=0}^{N-1} (M-i)}.
\end{eqnarray*}

\noindent
and $\alpha = q^{N-2} + $ h.o.t. One can also calculate the variance
of the first assembly time distribution to obtain

\begin{eqnarray*}
\nonumber
{\rm V}_N (M,0,\dots,0) &=&   \frac{\alpha^2}{\prod_{i=0}^{N-1}
  (M-i)^2} \\
&& - \frac{2 \beta}{\prod_{i=0}^{N-1}(M-i)}, 
\end{eqnarray*}

\noindent
and similarly all other moments of the distribution. Finally, we can
also estimate the first assembly time distribution $G(\{M,0
\dots,0\},t)$ by considering the inverse Laplace transform of
Eq.\,\ref{eq:laplace}, specifically by evaluating the dominant poles
associated to $d_{N-1}(s)$. In the large $q$ limit, $d_{N-1}(s)$ as
evaluated via the recursion relations Eqs.\,\ref{DISTR} can be
approximated as

\begin{eqnarray*}
d_{N-1}(s) \simeq q^{N-2} s +  \frac 1 2 \prod_{i=0}^{N-1} (M-i), 
\end{eqnarray*}

\noindent
yielding the slowest decaying root $\lambda_N$

\begin{eqnarray}
\label{eq65}
\lambda_{N} = - \frac 1 {2 q^{N-2}} \prod_{i=0}^{N-1} (M-i). 
\end{eqnarray}

\noindent
The above estimate allows us to write $G(\{M,0,\dots,0\}; t)$ in the
large $q$ limit as an exponential distribution with rate parameter
$\lambda_N$

\begin{eqnarray}
\label{eq66}
G(\{M,0,\dots,0\};t) \simeq \frac {e^{\lambda_N t}} 
{2 q^{N-2}} \prod_{i=0}^{N-1}
(M-i) 
\end{eqnarray}

\section{Hybrid approximation for $q\gg 1$ and $r < N-1$}
\label{APPENDIXD}

A more general hybrid approximation can be implemented by assuming
that only cluster of size $r$ and smaller pre-equilibrate.  We
integrate Eq.\,\ref{MASTER} over all configurations but with
$n_{r+1},...n_N$ fixed and obtain a reaction network for the remaining
$N-r$ clusters:

\begin{widetext}
\begin{equation}
\label{jackson_Network}
\begin{large}
\begin{array}{ccccccccc}
n_r & \xrightleftharpoons [q n_{r+1}]{ \langle n_1 n_r | \{n_{r+1}\}
\rangle_{\rm eq}} & n_{r+1} & \xrightleftharpoons[q n_{r+2}]{ \langle n_1 | \{
n_{r+1} \}\rangle_{\rm eq} n_{r+1}} & n_{r+2} & \cdots & \xrightarrow[]{
\langle n_1 | \{ n_{r+1} \} \rangle_{\rm eq} n_{N-1}} & n_N,
\end{array}
\end{large}
\end{equation}
\end{widetext}

\noindent
where $\{n_{r+1}\} = \{n_{r+1}, \cdots, n_N \}$ so that $\langle n_1
n_r | \{ n_{r+1} \} \rangle_{\rm eq}$ and $\langle n_1 | \{ n_{r+1} \}
\rangle_{\rm eq}$ depend on the slowly varying mass, $M - \sum_{r+1}^N
i n_i$, just as above for the choice $r = N-1$.

In the reaction chain (\ref{jackson_Network}) the last cluster
size $n_N$ is treated as an absorbing state since we are only
interested in the first assembly time, when $n_N=1$. The question
still remains of properly evaluating $\langle n_1 n_{N-1} |
\{n_{r+1}\} \rangle_{\rm eq}$ and $\langle n_1 | \{n_{r+1}\}
\rangle_{\rm eq}$.  For $q \to \infty$ it is reasonable to argue that
that most of the mass is distributed among the fast clusters
$n_1,...n_r$. Indeed, if we now assume that \textit{all} the mass is
contained in the fast cluster sizes, $\langle n_1 | \{n_{r+1}\}
\rangle_{\rm eq}$ and $ \langle n_1 n_{r} | \{ n_{r+1} \}
\rangle_{\rm eq}$ may be obtained via a distribution of $r$ clusters
with total mass $M - \sum_{i=r+1}^N i n_i \simeq M$.  The rates in
(\ref{jackson_Network}) become independent of $n_i$, for $i > r$. We
can also drop the slow cluster size condition on the averaged
quantities, and simply write $\langle n_1 \rangle_M$ and $\langle n_1
n_r \rangle_M$.

The cluster network in (\ref{jackson_Network}) is a so-called linear
Jackson queueing network \cite{Kingman1969}.  Entry of particles in
queue $n_{r+1}$ occurs at rate $\langle n_1 n_r \rangle_M$, each of
them moving independently according to the forward $ \langle n_1
\rangle_M$ and backward $q$ transition rates.  Starting with no
particles in the queue at $t=0$, the time-dependent probability
distribution for this queueing network is well known
\cite{Kingman1969}. In particular, the number of particles in the last
queue follows a Poisson distribution with mean

\begin{equation}
\mu(t)= \langle n_1 n_r \rangle_M \int_0^t {\cal P}_{N-r}(s) {\rm d}s
\nonumber
\end{equation}

\noindent
where ${\cal P}_{i}(t)$ is the probability that a single particle is in
the $i^{th}$ queue at time $t$ after its entry in the system. Because
the last queue is absorbing, and from the definition of the first
assembly time, the survival probability of our clustering process can
be identified with the probability of having no particles in the last
queue so that

\begin{eqnarray}
\label{start}
 S(t) &=& \mbox {Prob} \{n_N=0\} \nonumber \\ &=& 
 \exp\Big{[}- \langle n_1 n_r \rangle_M \int_0^t {\cal P}_{N-r}(s)
   \,{\rm d} s\Big{]} \nonumber.
\end{eqnarray}

\noindent
Finally, note the probability ${\cal P}_{i}(t)$, for $1\leq i \leq
N-r$ satisfies the Master equation $ \dot{\cal P}_{i}
=A_{ij} {\cal P}_{j}$ with $P_{i}(0) = \delta_{1i}$ and

\begin{widetext}
\begin{equation}
%\label{Matrix_Jackson}
A_{ij}= 
\left( \begin{array}{cccccc}
-q- \langle n_1 \rangle_M &\hspace{0.4cm} q & \hspace{0.4cm}  &  & & 0 \\
  \langle n_1 \rangle_M &\hspace{0.4cm} -q- \langle n_1 \rangle_M 
&\hspace{0.4cm} q & & & 0  \\
  &\hspace{0.4cm} \ddots &\hspace{0.4cm} \ddots & \ddots &  &0  \\
  &\hspace{0.4cm}  &\hspace{0.4cm}  &  \langle n_1 \rangle_M 
& -q- \langle n_1 \rangle_M & 0 \\
  &\hspace{0.4cm}  &\hspace{0.4cm}  & 0 &  \langle n_1 \rangle_M & 0 
\end{array} \right), \nonumber
\end{equation}
\end{widetext}

\noindent
The first assembly time and the variance can now be derived according
to standard formulae in Eq.\,\ref{timeone} and Eq.\,\ref{varone}.

As before, this technique requires an estimation of the first and
second moments $ \langle n_1 \rangle_M$ and $ \langle n_1 n_r
\rangle_M$ from the equilibrium distribution for clusters up to size $r$
with total mass $M$. A first crude way of approximating these asymptotic
moments is to use the mean-field results 

\begin{eqnarray*}
  \langle n_1 \rangle_M & \simeq & c_1^{\rm eq}, \quad
  \langle n_1 n_{N-1} \rangle_M  \simeq   c_1^{\rm eq} c_{N-1}^{\rm eq}.
\end{eqnarray*}

\noindent
We can also derive moment equations for $ \langle n_1 \rangle_M$ and $
\langle n_1 n_r \rangle_M$ directly from Eq.\,\ref{MASTER}. Here, due
to non-linear couplings between cluster sizes, the lower order moments
will necessarily be described in terms of higher order ones. For
instance, to determine the first and second moments we are interested
in, we would need an expression for the third moment.  To close moment
equations, one usually assumes that the probability distribution for
all cluster sizes obeys a certain form -- either Gaussian, log-normal
or negative binomial which are among the most standard. The third
moment may then be written as a function of the first two, thus
closing the system. The closed equations of the first two moments
become non-linear and a numerical solver is typically used to solve
them \cite{GILLESPIE}. The case $r=2$ has been extensively analyzed in
\cite{CAO}. In this paper we follow the same approach, using a
Gaussian distribution to approximate higher moments, thus deriving a
closed system of $r$ equations for $\langle n_i \rangle_M$ and
$r(r+1)/2$ equations for $\langle n_i n_j \rangle_M$, where $1 \leq
i,j \leq r$.  

%Alternatively, we may also run a numerical stochastic
%algorithm \cite{BKL75, GILLESPIE77} for sufficiently long times, and
%empirically compute the moments from the output of the numerical
%simulation \cite{ILIE,PINEDA}. We discuss all outcomes in the next
%section.

Finally, note that the hybrid approach described above is based on the
assumption that all mass is initially contained within the first $r$
clusters and are distributed according to the Becker-D\"oring
equilibrium distribution. We expect this approach to be valid for
moderate and large values of $M$ and $N$, with $q \geq M$ in order for
the production of small clusters to be faster than the production of
larger ones. How to choose the optimal cutoff value $r$ is a delicate
issue and depends on the specific parameters $\{M,N,q\}$, although in
general we find that all values of $ 2 \leq r \leq N-2$ give
qualitatively similar results.

%In the following section, we compare our 
%approximation methods with kinetic Monte Carlo simulations of the 
%self-assembly process. 

\end{appendix}

%%%%%%%%%%%%%%%%%%%%%%%%%%%%%%%%%%%%%%%%%%%%%%%%%%%%%%%%%%%%%%%%%%%
%%%%%%%%%%%%%%%%%%%%%%%%%%%%%%%%%%%%%%%%%%%%%%%%%%%%%%%%%%%%%%%%%%%
\end{document}